\documentclass[12pt]{article}

\usepackage{mathtools,amsthm,amssymb}
\usepackage{bm}
\usepackage{natbib}
\usepackage{xcolor}
\usepackage{multicol}
\usepackage{setspace}
\usepackage[ruled]{algorithm2e}
\usepackage{booktabs}
\usepackage{makecell}
\usepackage{graphicx}
\usepackage{caption}
\usepackage{authblk}
\usepackage{url}
\usepackage{hyperref}
\hypersetup{
  hypertexnames=false,
  colorlinks=true,
  allcolors=blue
}
\usepackage{bibunits}

\newtheorem{theorem}{Theorem}
\newtheorem{lemma}{Lemma}
\newtheorem{corollary}{Corollary}

\newtheorem{assumption}{Assumption}

\newcommand{\bA}{\mathbf{A}}

\newcommand{\bB}{\mathbf{B}}

\newcommand{\bc}{\mathbf{c}}

\newcommand{\bD}{\mathbf{D}}

\newcommand{\bI}{\mathbf{I}}

\newcommand{\bL}{\mathbf{L}}
\newcommand{\bM}{\mathbf{M}}
\newcommand{\bN}{\mathbf{N}}

\newcommand{\bQ}{\mathbf{Q}}

\newcommand{\bR}{\mathbf{R}}

\newcommand{\bbv}{\mathbf{v}}
\newcommand{\bW}{\mathbf{W}}

\newcommand{\bX}{\mathbf{X}}
\newcommand{\bx}{\mathbf{x}}

\newcommand{\by}{\mathbf{y}}

\newcommand{\cE}{\mathcal{E}}

\newcommand{\cN}{\mathcal{N}}

\newcommand{\bbeta}{\boldsymbol{\beta}}
\newcommand{\bdelta}{\boldsymbol{\delta}}

\newcommand{\bEta}{\boldsymbol{\eta}}

\newcommand{\bPi}{\boldsymbol{\Pi}}
\newcommand{\bmu}{\boldsymbol{\mu}}
\newcommand{\bnu}{\boldsymbol{\nu}}

\newcommand{\bSigma}{\boldsymbol{\Sigma}}

\newcommand{\btheta}{\boldsymbol{\theta}}

\newcommand{\bTheta}{\boldsymbol{\Theta}}

\newcommand{\E}{\mathrm{E}}

\newcommand{\rP}{\mathrm{P}}

\newcommand{\tr}{\mathrm{tr}}

\newcommand{\diag}{\mathrm{diag}}
\newcommand{\ber}{\mathrm{Bernoulli}}
\newcommand{\bzero}{\boldsymbol{0}}

\newcommand{\R}{\mathbb{R}}

\newcommand{\RN}[1]{
  \textup{\uppercase\expandafter{\romannumeral#1}}
}

\newcommand{\argmax}{\operatornamewithlimits{argmax}}
\newcommand{\argmin}{\operatornamewithlimits{argmin}}

\newenvironment{myproof}[1][\proofname]{
  \proof[\bfseries \upshape #1]
}{\endproof}

\def\spacingset#1{\renewcommand{\baselinestretch}
  {#1}\small\normalsize}

\date{}

\title{\bfseries Dynamic Supervised Principal Component Analysis for Classification}

\author[1]{Wenbo Ouyang}
\author[2]{Ruiyang Wu}
\author[1,3]{Ning Hao}
\author[1,3]{Hao Helen Zhang}

\affil[1]{GIDP in Statistics and Data Science, University of Arizona}
\affil[2]{Paul H. Chook Department of Information Systems and
  Statistics, Baruch College, The City University of New York}
\affil[3]{Department of Mathematics, University of Arizona}

\defaultbibliographystyle{chicago}
\defaultbibliography{DSPCAreference.bib}

\begin{document}

\maketitle
\spacingset{1}

\bigskip
\begin{abstract}
  This paper introduces a novel framework for dynamic classification
  in high dimensional spaces, addressing the evolving nature of class
  distributions over time or other index variables. Traditional
  discriminant analysis techniques are adapted to learn dynamic
  decision rules with respect to the index variable. In particular, we
  propose and study a new supervised dimension reduction method
  employing kernel smoothing to identify the optimal subspace, and
  provide a comprehensive examination of this approach for both linear
  discriminant analysis and quadratic discriminant analysis. We
  illustrate the effectiveness of the proposed methods through
  numerical simulations and real data examples. The results show
  considerable improvements in classification accuracy and
  computational efficiency. This work contributes to the field by
  offering a robust and adaptive solution to the challenges of
  scalability and non-staticity in high-dimensional data
  classification.
\end{abstract}

\vfill

\newpage
\spacingset{1.5} %

\begin{bibunit}
\section{Introduction}
\label{sec:intro}
Discriminant analysis is widely employed as a fundamental technique
for classification. In particular, Linear Discriminant Analysis (LDA)
strives to find a hyperplane that effectively separates data points
into different categories. LDA and its variants stand out as favorable
approaches to classification due to their simplicity and resilience in
handling the increasing dimensionality of contemporary datasets. In
addition, Quadratic Discriminant Analysis (QDA), which allows for data
heteroscedasticity, is another popular tool for nonlinear
classification. Recently, a variety of high-dimensional classifiers
have been proposed. These include sparse/regularized linear
classifiers
\citep{guo2007regularized,witten2011penalized,Shaoetal:2011,cai2011direct,fan2012road,mai2012direct},
dimension-reduction approaches
\citep{fan2008high,hao2015sparsifying,niu2015new}, and QDA-based
methods
\citep{li2015sparse,jiang2018direct,wu2019quadratic,wu2022quadratic},
to name just a few from a long list of references. The aforementioned
papers focus on only static models, assuming that the distribution of
each category remains unchanged throughout the data collection
process. However, this assumption may not be realistic in modern
applications, where the mean and covariance for each class might vary
over time or in response to an index variable. Consequently, there is
a demand for more adaptable and flexible modeling approaches to
account for such changes. To address this issue, dynamical modeling
has become popular in covariance estimation
\citep{yin2010nonparametric,chen2016dynamic,CHEN2019155,wang2021nonparametric},
classification \citep{jiang2020dynamic}, and principal component
analysis \citep{hu2024dynamic}. In particular, \cite{jiang2020dynamic}
proposed the Dynamic Linear Programming Discriminant (DLPD), which
accounts for the dynamic nature of the underlying data generation
mechanism. Unlike conventional static LDA, the DLPD approach is
capable of capturing the varying distribution of each underlying
population by modeling the mean and covariance as smooth functions of
an index variable. Moreover, the theoretical properties of the DLPD
are established under a high-dimensional and sparse setup.

In the literature of high-dimensional statistical learning, various
sparsity conditions have played important roles in modeling,
computing, and establishing theoretical guarantees
\citep{hastie2009elements}. In the context of classification, sparsity
conditions can be applied to mean differences
\citep{tibshirani2002diagnosis,fan2008high}, covariance
\citep{BickelLevina:2004}, or directly to the normal vector of the
discriminant hyperplane
\citep{cai2011direct,fan2012road,mai2012direct}. In particular, the
DLPD employs similar sparse assumptions as in~\cite{cai2011direct}.
Despite their widespread popularity, these sparsity conditions may not
align with the reality of many applications. It is beneficial to
explore alternative approaches under different model assumptions. For
instance, \cite{hao2015sparsifying} and~\cite{niu2015new} studied a
Supervised Principal Component Analysis (SPCA) approach for
high-dimensional classification. Instead of imposing explicit sparsity
conditions on the distribution parameters, they investigated an
implicit sparsity condition on the eigenvalues of the covariance and
provided a new classification strategy based on rotation and
projection. Furthermore, this approach demonstrates computational
efficiency in handling high-dimensional data, offering practitioners
more choices in data analysis.

In this article, we propose a flexible framework for high-dimensional
dynamic classification without explicit sparsity assumptions. Our
methodology is specifically designed to capture distributional changes
in dynamic contexts. The advantages of our method are fourfold. First,
it automatically learns a series of dynamic classification rules that
adapt to pattern changes in data over an index variable. Second, it
avoids explicit sparsity assumptions on the normal vector of the raw
data given the index variable; instead, our method effectively
discovers a suitable direction for data rotation to achieve sparsity.
Third, it is computationally efficient, without the requirement for
solving large-scale optimization problems. In addition, it essentially
serves as a dynamic dimension reduction tool that can be easily
combined with other classification methods. For example, we extend our
proposed framework to QDA, enabling us to address nonlinear
classification problems.

The rest of the paper is organized as follows. Section~\ref{sec2}
briefly reviews several modern classification tools relevant to our
approach. Section~\ref{sec3} proposes our new method called Dynamic
Supervised Principal Component Analysis. We illustrate numerical
experiments via simulated and real data examples in
Sections~\ref{sec4} and~\ref{sec5}, respectively. The proofs of the
theoretical results are given in the Supplementary Material.

We end this section with some notations used throughout this article.
For a vector \(\bbv = (v_j) \in \R^{p}\), define
\(\|\bbv\| = \sqrt{\bbv^{\top}\bbv}\), \(\|\bbv\|_1 = \sum_j |v_j|\)
and \(\|\bbv\|_{\infty} = \max_j |v_j|\). For a square matrix
\(\bM \in \R^{p \times p}\), let \(\tr(\bM)\) and
\(\lambda_{\max}(\bM)\) be its trace and greatest eigenvalue. For any
matrix \(\bM = (m_{jl}) \in \R^{p \times q}\), define
\(\|\bM\|_{F} = \sqrt{\tr(\bM^{\top}\bM)}\),
\(\|\bM\| = \sqrt{\lambda_{\max}(\bM^{\top}\bM)}\) and
\(\|\bM\|_{\infty} = \max_{j, l} |m_{jl}|\). For real numbers \(a\)
and \(b\), let \(a \wedge b = \min\{a, b\}\) and
\(a \vee b = \max\{a, b\}\). For sequences of real numbers \((a_n)\)
and \((b_n)\), we write \(a_n \lesssim b_n\) or \(a_n = O(b_n)\) if
there exists some constant \(C > 0\) such that \(|a_n| \leq C|b_n|\)
for all sufficiently large \(n\). Let \(a_n \asymp b_n\) denote
\(a_n \lesssim b_n\) and \(b_n \lesssim a_n\). Finally, we use \(D^l\)
for the \(l\)th order derivative operator.

\section{Discriminant Analysis in High Dimensions}
\label{sec2}

\subsection{Linear Discriminant Analysis and Its Variants}

Let $\bX\in{\R^p}$ be a random vector, and $Y \in\{1,2\}$ be its class
label. Assuming the conditional distribution
$\bX|\{Y=c\}\sim \mathcal{N}(\bmu^{(c)},\bSigma)$ and the prior
probability $\pi_{c}=\rP(Y=c)$ for $c\in\{1,2\}$, one can derive the
optimal classification rule, also called the Bayes rule, which labels
an observation ${\bx}\in{\R^p}$ based on the sign of the discriminant
function
\begin{eqnarray*}
f(\bx)=({\bx}-\bmu)^{\top}\bSigma^{-1}\bdelta+\text{log}(\pi_{1}/\pi_{2})=(\bx-\bmu)^{\top}\bbeta+\text{log}(\pi_{1}/\pi_{2}),
\end{eqnarray*}
where
$\bdelta=\bmu^{(1)}-\bmu^{(2)},
\bmu=\frac{1}{2}(\bmu^{(1)}+\bmu^{(2)})$, and
$\bbeta=\bSigma^{-1}\bdelta$. The hyperplane defined by $f(\bx)=0$ is
called the optimal decision boundary. In general, LDA and its variants
aim to approximate the optimal decision boundary using the training
data. For example, in the standard LDA, the normal vector $\bbeta$ is
usually estimated by plugging empirical means and covariance in the
formula above, when the sample size $n$ is larger than $p$. That is,
the standard LDA labels data according to
\begin{eqnarray*}
\hat f(\bx)=({\bx}-\hat\bmu)^{\top}\hat\bSigma^{-1}\hat\bdelta+\text{log}(\hat\pi_{1}/\hat\pi_{2})=(\bx-\hat\bmu)^{\top}\hat\bbeta+\text{log}(\hat\pi_{1}/\hat\pi_{2}),
\end{eqnarray*}
where $\hat\bbeta=\hat\bSigma^{-1}\hat\bdelta$,
$\hat\bdelta=\hat \bmu^{(1)}-\hat \bmu^{(2)}$,
$\hat\bmu=\frac{1}{2}(\hat{\bmu}^{(1)}+\hat{\bmu}^{(2)})$,
$\hat\bmu^{(1)}$ and $\hat \bmu^{(2)}$ are the sample means of the two
classes, $\hat\pi_{1}$ and $\hat\pi_{2}$ are the sample proportions,
and $\hat\bSigma$ is the pooled sample covariance.

Fisher's Discriminant Analysis \citep{fisher1936use} aims to identify
a direction that maximizes the separation between the projected class
means relative to their individual spread, which can be formulated as:
\begin{eqnarray}
  \label{eq:1}
  \hat\bbv=\argmax_{ \|\bbv\|=1} \frac{\bbv^{\top}\hat\bdelta\hat\bdelta^{\top}\bbv}{\bbv^{\top}\boldsymbol{\hat\Sigma}\bbv}.
\end{eqnarray}
It is straightforward to show $\hat\bbeta=c\hat\bbv$ for a nonzero
scalar $c$. Besides offering a geometric interpretation of LDA,
Fisher's approach inspires many high-dimensional LDA methods. To
elaborate, first, observe that~\eqref{eq:1} is equivalent to the
optimization
\begin{eqnarray}
  \label{eq:2}
  \hat\bbv=\argmin_{\bbv: \|\bbv^{\top}\hat\bdelta\|=1}  {\bbv^{\top}{\hat\bSigma}\bbv}.
\end{eqnarray}
This formulation can be easily adapted to modern regularization
frameworks. For example, a few popular high-dimensional classification
tools such as~\cite{wu2009sparse} and~\cite{fan2012road} employ
regularized versions of~\eqref{eq:2}:
\begin{eqnarray*}
\min_{\bbeta} \bbeta^{\top}\hat\bSigma\bbeta \text{\quad subject to } \;  \|\bbeta^{\top}\hat\bdelta\|=1,\, \|\bbeta\|_1\leq C,
\end{eqnarray*}
and
\begin{eqnarray*}
\min_{\bbeta}\bbeta^{\top}\hat\bSigma\bbeta +\lambda \|\bbeta\|_1\text{,\quad subject to } \;  \|\bbeta^{\top}\hat\bdelta\|=1,
\end{eqnarray*}
where $C$ and $\lambda$ are tuning parameters controlling the sparsity
of the solutions\@. \cite{mai2012direct} converts the LDA problem as a
linear regression problem and then performs variable selection by the
LASSO \citep{tibshirani1996regression}. Later, \cite{mai2013note}
shows the equivalence of these regularized LDA approaches\@.
\cite{cai2011direct} utilizes a different regularization framework
called Linear Programming Discriminant (LPD), which solves
\begin{eqnarray*}
\min_{\bbeta} \|\bbeta\|_1\text{,\quad subject to } \;  \|\hat\bSigma\bbeta-\hat\bdelta\|_{\infty}\leq \lambda.
\end{eqnarray*}

The aforementioned classification tools are widely used in various
applications. It is important to note that the performance of these
regularized LDA methods is, to a considerable degree, contingent upon
the sparsity level of the normal vector $\bbeta$.

\subsection{Supervised Principal Component Analysis}
Instead of directly estimating the normal vector $\bbeta$ under
sparsity conditions, an alternative strategy involves approximating a
lower-dimensional subspace that contains $\bbeta$, followed by the
estimation of $\bbeta$ using conventional methods such as the standard
LDA estimator. Such subspaces are feasible under a spiked condition on
the covariance, as explored in \citep{hao2015sparsifying,niu2015new}.
Specifically, \cite{niu2015new} proposes a Supervised Principal
Component Analysis (SPCA) approach for conducting dimension reduction.
To elaborate, define a total covariance matrix $\bSigma_{\rho}^{tot}$
as a weighted sum of the within-class covariance and the between-class
covariance:
\begin{eqnarray*}
  \bSigma_{\rho}^{tot}=\bSigma+\rho\bdelta \bdelta^{\top}\text{,\quad where} \;  \rho>0.
\end{eqnarray*}

SPCA employs the top $K$ eigenvectors of $\bSigma_{\rho}^{tot}$ for
dimension reduction and classification. This method depends on two
tuning parameters $\rho$ and $K$. Consider the eigen-decomposition
\begin{eqnarray*}
  \bD_{\rho}=\bR^{\top}_{\rho}\bSigma_{\rho}^{tot}\bR_{\rho},
\end{eqnarray*}
where $\bD_{\rho}$ is a diagonal matrix with eigenvalues listed in a
descending order, and $\bR_{\rho}$ is an orthogonal matrix. If the
common covariance matrix $\bSigma$ is spiked, i.e., all of its
eigenvalues are the same except for $s$ larger ones, then the normal
vector $\bbeta$ to the optimal discriminant boundary is located in the
linear subspace spanned by first $s+1$ eigenvectors of
$\bSigma_{\rho}^{tot}$, represented by the left $s+1$ columns of the
matrix $\bR_{\rho}$. Consequently, we can project the data to this
$(s+1)$-dimensional subspace without losing discriminant power. In
practice, an empirical version of the total covariance is used to
approximate the subspace. The parameters $\rho$ and $K$ are often
selected by cross-validation. In practice, the SPCA approach to
classification performs well even if the spiked condition is not
satisfied. In summary, this SPCA method offers an alternative and
competitive way for high-dimensional classification.

\section{Dynamic Supervised Principal Component Analysis}
\label{sec3}

\subsection{Dynamic Discriminant Analysis}
\label{sec:dynam-discr-analys}

Consider the binary classification problem in a dynamic scenario,
where the distributions of the observations may change dynamically
with respect to an index variable $U$. Given the training data
consisting of independent observations $(\bx_i, u_i, y_i)$ for
$i=1, \ldots, n$, where $\bx_i\in \R^p$, $u_i\in \R$, $y_i\in\{1,2\}$,
the goal is to learn a classification rule and predict the label for a
new observation $(\bx,u)$. Consider a dynamic LDA model
$\bX|{(Y=c, U=u)} \sim \mathcal{N}(\bmu^{(c)}(u),\bSigma(u))$, where
$\bmu^{(c)}(u)$ is the $p$-dimensional mean vector for $c = 1, 2$, and
$\bSigma(u)$ is the $p\times p$ common covariance matrix. Both the
covariance and mean vectors are functions of the index variable $U$.
Consequently, the optimal decision boundary depends on $U$. In this
situation, it is suboptimal to apply a static classification tool.
It is challenging to estimate
the normal vector $\bbeta(U)$ of the optimal discriminant
hyperplane\@. \cite{jiang2020dynamic} proposed the Dynamic Linear
Programming Discriminant (DLPD), which estimates the means and
covariance as smooth functions of an index variable $U$ and then
employs the LPD rule for classification. In spite of the theoretical
guarantee shown for DLPD, its expensive computation cost makes it less
appealing in applications involving high-dimensional data.

We next propose a new method called Dynamic Supervised Principal
Component Analysis (DSPCA) to learn classification decision rules that
vary with the index $U$. Recall that, conditional on the label $Y=c$
and the index variable $U=u$, $\bX$ is of mean $\bmu^{(c)}(u)$, and
variance $\bSigma(u)$. We define the parameters as well as the total
covariance as functions of the index variable:
\begin{eqnarray}
    \bdelta(u)&=&\bmu^{(1)}(u)-\bmu^{(2)}(u),\nonumber \\
    \bmu(u)&=&\frac{1}{2}\left(\bmu^{(1)}(u)+\bmu^{(2)}(u)\right),\nonumber\\
    \bSigma_{\rho}^{tot}(u)&=&\bSigma(u)+\rho\bdelta(u)\bdelta^{\top}(u),\quad \rho>0.\label{eq:3}
\end{eqnarray}
The DSPCA conducts dimension reduction based on top eigenvectors of
the total covariance $\bSigma_{\rho}^{tot}(u)$ when $U=u$. A simple
classification tool such as the standard LDA can then be applied on
the reduced space. To elucidate, we first diagonalize
$\bSigma_{\rho}^{tot}(u)$ as
\begin{eqnarray}
  \label{eq:4}
  \bR^{\top}(u)\bSigma_{\rho}^{tot}(u)\bR(u)=\bD(u),
\end{eqnarray}
where $\bD(u)$ is a diagonal matrix, and $\bR(u)$ is an orthogonal
matrix formed by the eigenvectors corresponding to the sorted
eigenvalues of $\bSigma_{\rho}^{tot}(u)$. Then we define a new feature
vector $\tilde\bx_{i}$ by the rotation
$\tilde\bx_{i}=\bR^{\top}(u)\bx_{i}, i=1, \ldots, n$. $\tilde\bx_{i}$
is simply the new coordinate if we change the basis of the feature
space using the eigenvectors of the total covariance. In
Section~\ref{sec3.5}, we will demonstrate why only the first several
coordinates are critical for classification. Consequently, we propose
employing the standard LDA method on these foremost coordinates of the
rotated data. In practical scenarios, the unknown parameters
$\bmu^{(c)}(u)$ and $\bSigma(u)$ are estimated by the kernel smoothing
method. The whole procedure depends on three tuning parameters: the
bandwidth $h$ in kernel smoothing, the weight $\rho$ in the total
covariance, and the dimension $K$ of the reduced subspace. We will
address related nonparametric estimation, tuning parameter selection,
and computation issues in the next several subsections.

We end this subsection with a brief remark. The classical PCA can be
considered as a special case of SPCA, which can balance the estimated
within-class and between-class covariances via the parameter $\rho$
and achieve better classification accuracy in the reduced space
\citep{niu2015new}. Both PCA and SPCA are static methods, while our
new proposal generalizes SPCA to the dynamic situation.

\subsection{Parameter Estimation}

In practice, we need to estimate the unknown model parameters
$\bmu^{(c)}(u)$ and $\bSigma(u)$ using the training data. One natural
and effective approach for this task is the Nadaraya-Watson estimator,
which leverages a kernel function to construct a locally weighted
average sample estimator. The Nadaraya-Watson estimator assigns higher
weights to observations that are closer to the target point, making it
well-suited for dynamic settings where both the means and the
covariance matrix can vary across different target points as a
function of $u$. An advantage of the Nadaraya-Watson estimator is its
robustness to model misspecifications. As a nonparametric estimation
procedure, it relies on data-driven principles rather than making
assumptions about the specific form of the function that characterizes
the relationship between the estimator and the index variable $u$.
This data-driven nature allows the Nadaraya-Watson estimator to adapt
flexibly to various underlying structures, making it a valuable tool
for estimating model parameters without being bound by specific
assumptions. Other nonparametric smoothing methods, such as local
polynomial regression and smoothing splines, share similar advantages
and are also applicable for parameter estimation. In this paper, we
focus on the Nadaraya-Watson estimator for its simplicity in both
theory and computation.

Let $\{(\bx_{i},u_i,y_i): 1\le i\le n\}$ be a random sample. The
kernel function with a bandwidth $h$ is defined as
\begin{eqnarray*}
K_{h}(u)=\frac{1}{h}K\left(\frac{u}{h}\right),
\end{eqnarray*}
where $K(\cdot)$ is a univariate density function, e.g., the standard
Gaussian density function. The Nadaraya-Watson estimator for the mean
vector is
\begin{equation*} 
\hat{\bmu}^{(c)}(u)=\frac{\sum_{i: y_i=c }K_h(u_{i} -u)\bx_i}{\sum_{i: y_i=c }K_h(u_{i}-u)}, \quad c=1,2
\end{equation*}
where the bandwidth $h$ parameter is chosen adaptively, e.g., by
leave-one-out cross-validation. The kernel estimators for the
covariance matrix of each class is
\begin{multline*} 
\hat{\bSigma}^{(c)}(u)=\frac{\sum_{i: y_i=c }K_{h}(u_{i}-u)\bx_{i}\bx_{i}^{\top}}{{\sum_{i: y_i=c }K_{h}(u_{i}-u)}} \\-\frac{\left(\sum_{i: y_i=c }K_{h}(u_{i}-u)\bx_{i}\right)\left(\sum_{i: y_i=c }K_{h}(u_{i}-u) \bx_{i}^{\top}\right)}{\left({\sum_{i: y_i=c }K_{h}(u_{i}-u)}\right)^{2}},\; c=1,2.
\end{multline*}
Then, the pooled estimator of the covariance matrix is given by the
weighted average of two kernel estimators
\begin{eqnarray*} 
\hat{\bSigma}(u)=\frac{n_1}{n}\hat{\bSigma}^{(1)}(u)+\frac{n_{2}}{n}\hat{\bSigma}^{(2)}(u),
\end{eqnarray*}
where $n_1$ and $n_2$ are the numbers of observations in classes 1 and
2, respectively.

\subsection{Implementation}
\label{sec:implementation}

Based on the estimators $\hat{\bmu}^{(1)}(u)$, $\hat{\bmu}^{(2)}(u)$,
and $\hat{\bSigma}(u)$, we define
$\hat\bdelta^{\top}(u)=\hat{\bmu}^{(1)}(u)-\hat{\bmu}^{(2)}(u)$ and
\begin{eqnarray}
  \label{eq:5}
  \hat{\bSigma}_{\rho}^{tot}(u)=\hat{\bSigma}(u)+\rho\hat{\bdelta}(u) \hat\bdelta^{\top}(u),\quad\rho>0.
\end{eqnarray}
In the eigen-decomposition
$\hat{\bR}^{\top}(u)\hat{\bSigma}_{\rho}^{tot}(u)\hat{\bR}(u)=\hat{\bD}(u)$,
let $\hat{\bD}(u)=\mbox{diag}(\lambda_1(u), \dots, \lambda_{p}(u))$ be
a diagonal matrix with $\lambda_1(u)\geq \cdots\geq \lambda_{p}(u)$,
and $\hat{\bR}(u)$ be an orthogonal matrix formed by the eigenvectors
of $\hat{\bSigma}_{\rho}^{tot}(u)$. For a target dimension $K$, we
write $\hat{\bR}(u)=(\hat{\bR}_1(u),\hat{\bR}_2(u))$, where
$\hat{\bR}_1(u)$ is a $p\times K$ matrix and $\hat{\bR}_2(u)$ is a
$p\times (p-K)$ matrix. Note that the columns of $\hat{\bR}_1(u)$ are
eigenvectors of $\hat{\bSigma}_{\rho}^{tot}(u)$ corresponding to top
$K$ eigenvalues. Therefore, $\hat{\bR}_1^{\top}(u)\bx_{i}$ is the
projected vector. Typically we will choose a $K$ smaller than $n$, so
the standard LDA can be applied to the projected data. In summary, the
proposed DSPCA methodology comprises three steps. Initially, a kernel
method is utilized to construct nonparametric estimators to the mean
vectors and covariance from the training sample. Subsequently,
dimension reduction is carried out by a modified SPCA procedure
incorporating the nonparametric estimators. Lastly, the LDA rule is
applied to the projected data to derive the decision rule. The
pseudo-code detailing the implementation of the entire procedure is
presented in Algorithm~\ref{alg:1}.

\begin{algorithm}[!htbp]
  \spacingset{1.5}

  \KwIn{Training data $\{(\bx_{i},u_{i},y_i): 1\le i\le n\}$, tuning
    parameters $\rho$, $K$, $h$, an unlabeled observation
    $(\bx^{*}, u^{*}$).}

  \KwOut{The predicted label $\hat y$ for \((\bx^*, u^*)\).}

  \textbf{Step 1:} Use all the training data to calculate the
  Nadaraya-Watson estimator
  $\hat{\bmu}^{(1)}(u^{*}),\hat{\bmu}^{(2)}(u^{*}),
  \hat{\bSigma}^{(1)}(u^{*})$, and $\hat{\bSigma}^{(2)}(u^{*})$.

  \textbf{Step 2a:} Compute $\hat{\bSigma}_{\rho}^{tot}(u^{*})$ as
  in~\eqref{eq:5}.

  \textbf{Step 2b:} Apply spectral decomposition to
  $\hat{\bSigma}_{\rho}^{tot}(u^{*})$ and take top $K$ eigenvectors.

  \textbf{Step 3:} Project the training data and $\bx^{*}$ to the
  linear subspace spanned by the $K$ vectors in Step 2b, then apply
  the LDA rule to the projected data for classification.

  \caption{Dynamic Supervised Principal Component Analysis (DSPCA)}
  \label{alg:1}
\end{algorithm}

For high dimensional data where $p$ is much larger than the sample
size $n$, it is time-consuming to conduct the spectral decomposition
to the $p\times p$ matrix $\hat{\bSigma}_{\rho}^{tot}(u)$. We use a
trick to speed up the computation. Note that we can decompose
$\hat{\bSigma}_{\rho}^{tot}(u)$ as
$\bA_{\rho}(u)^{\top}\bA_{\rho}(u)$, where $\bA_{\rho}(u)$ is a
$(n+1) \times p$ matrix by Lemma 1 in~\cite{niu2015new}. It is much
faster to conduct spectral decomposition to the $(n+1) \times (n+1)$
matrix $\bA_{\rho}(u)\bA_{\rho}(u)^{\top}$, which shares the same
nonzero eigenvalues with $\hat{\bSigma}_{\rho}^{tot}(u)$. Moreover,
all the eigenvectors (corresponding to the nonzero eigenvalues) of
$\hat{\bSigma}_{\rho}^{tot}(u)$ can be obtained from eigenvectors of
$\bA_{\rho}(u)\bA_{\rho}(u)^{\top}$ through a linear transformation
$\bA_{\rho}(u)^{\top}$. This trick is particularly useful for dealing
with high-dimensional data such as gene expressions. %

\subsection{Tuning}

The DSPCA method proposed in this study involves several tuning
parameters including $h$, $\rho$, and $K$, which need to be chosen
data-adaptively to achieve optimal performance in practical
applications. For selecting the bandwidth $h$ in the kernel estimator,
we employ the leave-one-out cross-validation procedure. Specifically,
we use $\hat{\bmu}_{-i}^{(c)}(u_{i})$ and
$\hat{\bSigma}_{-i}^{(c)}(u_{i})$ to represent the estimates of the
mean and covariance at $u_{i}$, respectively, which are obtained from
the Nadaraya-Watson estimator using all observations in class $c$,
except for the $i$th observation. To choose the best $h$ for
estimating $\bmu^{(c)}(u)$, we consider a wide range of potential
values and select the one that minimizes \(Err_{mean.cv}^{(c)}(h)\) as
defined in~\eqref{eq:6} below. Similarly, the optimal bandwidth for
covariance estimation is chosen using \(Err_{var.cv}^{(c)}(h)\)
in~\eqref{eq:7} as the criterion. Note that each class may have
different bandwidths tuned independently with the procedure outlined
above, and all the bandwidths are determined before conducting
dimension reduction.
\begin{align}
Err_{mean.cv}^{(c)}(h)&=\frac{1}{p^{2}n_{c}} \sum_{i: y_i=c}(\bx_{i}-\hat{\bmu}_{-i}^{(c)}(u_{i}))^{\top}(\bx_{i}-\hat{\bmu}_{-i}^{(c)}(u_{i})),\label{eq:6}\\
Err_{var.cv}^{(c)}(h)&=\frac{1}{p^{2}n_{c}} \sum_{i: y_i=c}\left\|(\bx_{i}-\hat{\bmu}^{(c)}_{-i}(u_{i}))(\bx_{i}-\hat{\bmu}^{(c)}_{-i}(u_{i}))^{\top}-\hat{\bSigma}^{(c)}_{-i}(u_{i})\right\|^{2}_{F}.\label{eq:7}
\end{align}

For the dimension reduction procedure, we employ 5-fold
cross-validation to select the tuning parameters, i.e., $\rho$ as in
the definition~\eqref{eq:5} of total covariance, and the dimension $K$
of the reduced space. We found that the performance of DSPCA is not
sensitive to the choice of $\rho$ when it varies in a small range, so
we suggest choosing a few $\rho$ from a relatively big range. Our
default range of choices for $\rho$ is the set
$\mathfrak{R}=\{{\exp(\mathfrak{r}): \mathfrak{r}=-1,0,\dots, 6}\}$.
For each $\rho$, we find the top $K_{\max}$ eigenvectors. Then we
project the data onto the $K_{\max}$-dimensional reduced space and
apply the LDA rule to the first $K$ coordinates, where $K$ takes
integer values in $\mathfrak{K}=\{1,\dots,K_{\max}\}$. The choice of
$K_{\max}$ may depend on the sample size and any prior knowledge of
the covariance structure. $K_{\max}=5$ is used in our numerical
studies. Finally, within the grid $\mathfrak{R} \times \mathfrak{K}$,
the combination of $\rho$ and $K$ that minimizes the cross-validation
classification error is chosen to conduct DSPCA\@. When such a
minimizer is not unique, we first find and fix the smallest admissible
$K$ value, then take the smallest corresponding $\rho$ as our choice.

\subsection{Theoretical Results}
\label{sec3.5}
In high-dimensional contexts, estimating the population eigenvalues
and eigenvectors from the sample covariance matrix often proves
challenging due to ill-conditioning and numerical complexities. To
address these challenges, the spiked covariance model
\citep{johnstone2001distribution} is often employed to depict the
covariance structure of high-dimensional data. This model posits that
eigenvectors corresponding to spiked eigenvalues can be consistently
estimated under certain conditions. Consequently, this study considers
the problem of dimension reduction within a spiked covariance
framework. Let $\{\lambda_{j}(u)\}^{p}_{j=1}$ denote the set of
eigenvalues of $\bSigma(u)$ with
$\lambda_{1}(u)\geq\lambda_{2}(u)\geq\cdots\geq\lambda_{p}(u)$. A
spiked structure on the covariance $\bSigma(u)$ assumes that all the
eigenvalues are equal except for the top $k$ eigenvalues, i.e.,
$\lambda_1(u),\ldots,\lambda_k(u)$, where $k$ is usually assumed to be
much smaller than $p$. %
The following Theorem~\ref{thm:1} characterizes a linear subspace that
contains the normal vector \(\bbeta(u) = \bSigma(u)^{-1}\bdelta(u)\)
of the optimal discriminant boundary.

\begin{theorem}
  \label{thm:1}
  Assume a dynamic LDA model
  $\bX|{(Y=c, U=u)} \sim \mathcal{N}(\bmu^{(c)}(u),\bSigma(u))$,
  $c=1,2$, where the eigenvalues of $\bSigma(u)$ satisfy
  $\lambda_{1}(u)\geq\cdots\geq\lambda_{k}(u)>\lambda_{k+1}(u)=\cdots=\lambda_{p}(u)$,
  for an integer $k$. In the eigen-decomposition~\eqref{eq:4}, we
  write $\bR(u)=(\bR_1(u),\bR_2(u))$ where $\bR_1(u)$ and $\bR_2(u)$
  are $p\times(k+1)$ and $p\times (p-k-1)$ matrices, respectively.
  Then we have $\bR_2(u)^{\top}\bbeta(u) =0$. In other words, the
  $\bbeta(u)$ is located in the linear subspace spanned by columns of
  $\bR_1(u)$.
\end{theorem}

Theorem 1 implies that all information about the class label \(Y\) is
carried only in the first \(k+1\) coordinates of the rotated data
\(\bR(U)^{\top}\bX\). This elucidates the efficacy of our DSPCA
approach for classification, based on the premise that
\(\bR_1(u^*)^{\top}\bx^*\) can be well estimated for any given
unlabeled test data \((\bx^*, u^*)\). Consequently, the success of
DSPCA primarily rests on effective control of the distance between
\(\bR_1(u^*)\) and \(\hat \bR_1(u^*)\), where \(\hat \bR_1(u^*)\)
consists of the first \(k + 1\) columns of \(\hat \bR(u^*)\). For the
remainder of this section, we aim to derive a uniform upper bound of
\[d(\bR_1(u), \hat\bR_1(u)) \coloneq \|\bR_1(u)\bR_1(u)^{\top} -
  \hat\bR_1(u)\hat\bR_1(u)^{\top}\|\] for all possible values of the
index variable \(U\).

To obtain the main result, we make some necessary assumptions. First
note that the spiked covariance model assumption allows \(\bSigma(u)\)
to be decomposed as
\begin{align}
  \bSigma(u) &\overset{\mathrm{(i)}}{=} \bQ(u)\diag(\lambda_1(u), \dots, \lambda_p(u))\bQ(u)^{\top}\nonumber\\
             &= \bQ(u)\diag(\lambda_1(u) - \sigma(u)^2, \dots, \lambda_k(u) - \sigma(u)^2, 0, \dots, 0)\bQ(u)^{\top} + \sigma(u)^2\bI_p\nonumber\\
             &= \bQ_1(u)\diag(\lambda_1(u) - \sigma(u)^2, \dots, \lambda_k(u) - \sigma(u)^2)\bQ_1(u)^{\top} + \sigma(u)^2\bI_p\nonumber\\
             &= \bL(u)\bL(u)^{\top} + \sigma(u)^2\bI_p,\label{eq:8}
\end{align}
where (i) is the eigen-decomposition of \(\bSigma(u)\), \(\sigma(u)\)
is defined as
\(\sigma(u) = \sqrt{\lambda_{k + 1}(u)} = \cdots =
\sqrt{\lambda_{p}(u)}\), \(\bQ_1(u)\) is a \(p \times k\) matrix
containing the first \(k\) columns of \(\bQ(u)\), and \(\bL(u)\) is a
\(p \times k\) matrix defined by
\(\bL(u) = \bQ_1(u)\diag(\sqrt{\lambda_1(u) - \sigma(u)^2}, \dots,
\sqrt{\lambda_k(u) - \sigma(u)^2})\).

\begin{assumption}
  \label{asp:1}
  \(f_U\), the probability density function of \(U\), satisfies: 1.\
  \(f_U = 0\) outside \([0, 1]\), 2.\ \(f_U \geq C_U\) on \([0, 1]\),
  and 3.\ \(f_U\) is twice continuously differentiable on \([0, 1]\)
  with
  \(\max_{l = 0, 1, 2}\sup_{u \in [0, 1]}|D^l f_U(u)| \leq \tilde
  C_U\), where \(C_U, \tilde C_U > 0\) are universal constants.
\end{assumption}
\begin{assumption}
  \label{asp:2}
  \(\bmu^{(1)}(u)\) and \(\bmu^{(2)}(u)\) are twice continuously
  differentiable on \([0, 1]\). Define
  \(M \coloneq \max_{c, l = 1, 2}\sup_{u \in [0, 1]}\|\bmu^{(c)}(u)\|
  \vee \|D^{l}\bmu^{(c)}(u)\|_{\infty}\).
\end{assumption}
\begin{assumption}
  \label{asp:3}
  \(\bL(u)\) and \(\sigma(u)\) are twice continuously differentiable
  on \([0, 1]\). Define
  \(\gamma \coloneq \max_{l = 0, 1, 2} \sup_{u \in [0, 1]} |D^l
  \sigma(u)|\),
  \(\Delta_1 \coloneq \max_{l = 1, 2} \sup_{u \in [0, 1]}\|\bL(u)\|
  \vee \|D^l \bL(u)\|_{\infty}\), and
  \(\Delta_k \coloneq \inf_{u \in [0, 1]}\sqrt{\lambda_k(u) -
    \sigma(u)^2}\).
\end{assumption}
\begin{assumption}
  \label{asp:4}
  There exists a linear subspace \(W \subseteq \R^p\) with
  \(\dim(W) = r\), such that for any \(u \in [0, 1]\), the linear
  subspace spanned by the columns of \(\bL(u)\) is contained in \(W\).
\end{assumption}
\begin{assumption}
  \label{asp:5}
  \(\sin^2(\phi) \geq C_{\phi}\), where \(\phi\) is the angle between
  \(\bdelta(u)\) and the column space of \(\bL(u)\), and
  \(C_{\phi} > 0\) is a universal constant.
\end{assumption}
\begin{assumption}
  \label{asp:6}
  The kernel function \(K(\cdot)\) satisfies: 1.\ \(K(u) = K(-u)\),
  2.\ \(\int_{\R} K(u)du = 1\), 3.\
  \(\max_{l = 1, 2, 3}\int_{\R} u^{2l}K(u)du \leq C_K\), 4.\
  \(\int_{\R}K^2(u)du \leq C_K\), and 5.\
  \( \max_{l = 0, 1}\sup_{u\in \R}|D^{l}K(u)| \leq C_K\), where
  \(C_K > 0\) is a universal constant.
\end{assumption}

Assumptions~\ref{asp:1}--\ref{asp:3} postulate the smoothness of the
density function of the index variable \(U\), as well as that of the
mean and covariance functions. We assume the domain of \(U\) to be the
unit interval \([0, 1]\) for concision and clarity, which can be
generalized to any finite closed interval of \(\R\). The
decomposition~\eqref{eq:8} implies that the covariance can be roughly
represented as the sum of a low-rank component,
\(\bL(u)\bL(u)^{\top}\), and a spherical component,
\(\sigma(u)^2\bI_p\), for a fixed \(u\). Assumption~\ref{asp:4}
ensures that the low-rank property of \(\bL(u)\bL(u)^{\top}\) holds
when \(u\) varies. Assumption~\ref{asp:5} is a non-essential technical
condition which excludes a singular case where both the mean
difference \(\bdelta(u)\) and the normal vector \(\bbeta(u)\) lie in
the subspace spanned by the first \(k\) spiked eigenvectors of
\(\bSigma(u)\). In this scenario, the total covariance
\(\hat\bSigma^{tot}_{\rho}(u)\) gains higher signal-to-noise ratio
than \(\hat\bSigma(u)\) because the added term
\(\rho\hat\bdelta(u)\hat\bdelta(u)^{\top}\) mainly inflates the
variance along the first \(k\) principal components, giving an
advantage to our algorithm over the classical PCA performed on the
within-class covariance. Particularly, Assumption~\ref{asp:5}
prohibits the trivial case where the two classes are the same, i.e.,
\(\bdelta(u) = \bzero\). This allows us to define
\(m \coloneq \sqrt{\rho}\inf_{u \in [0, 1]}\|\bdelta(u)\| > 0\).
Assumption~\ref{asp:6} is a standard condition in the kernel smoothing
literature \citep{hu2024dynamic,jiang2020dynamic,Pagan_Ullah_1999}.

\begin{theorem}
  \label{thm:2}
  Suppose that \(\log p/(hn) \to 0\), \(h \to 0\), \(\rho = O(1)\) and
  \(B_n = o(\Delta_k^2 \wedge m^2)\), where \(B_n > 0\) and its
  definition is deferred to the Appendix. Under
  Assumptions~\ref{asp:1}--\ref{asp:6}, we have as
  \(n, p \to \infty\),
  \[\sup_{u \in [0, 1]}d(\bR_1(u), \hat \bR_1(u)) \lesssim
    B_n (\Delta_k^2 \wedge m^2)^{-1}\] with probability larger than
  \(1 - O(h^{-4}np^{-11.5})\).
\end{theorem}

Theorem~\ref{thm:2} establishes the performance baseline of our DSPCA
algorithm with a convergence rate of
\(B_n (\Delta_k^2 \wedge m^2)^{-1}\). It guarantees consistent
estimation of the principal components for a broad class of LDA
models. The following Corollary~\ref{cor:1} specifies such a model
class with mild conditions on its parameters. Notably, the dimension
\(p\) is allowed to grow much faster than \(n\) and only slightly
slower than \(n^2\), even though we do not impose explicit sparsity
assumptions on the normal vector \(\bbeta(u)\).
\begin{corollary}
  \label{cor:1}
  In addition to Assumptions~\ref{asp:1}--\ref{asp:6}, suppose that
  \begin{enumerate}
  \item \(p^{\frac{1}{5}}(\log p)^2 \lesssim n^{\frac{2}{5}}\),
  \item \(\Delta_1^2 \vee M^2 \lesssim \Delta_k^2 \wedge m^2\),
    \(\gamma^2 \lesssim (\Delta_k^2 \wedge m^2)p^{-1}\),
  \item \(h \asymp (\log p/(p^2n))^{\frac{1}{5}}\),
  \item \(\rho, r, k = O(1)\).
  \end{enumerate}
  Then we have as \(n, p \to \infty\),
  \[\sup_{u \in [0, 1]}d(\bR_1(u), \hat \bR_1(u)) \lesssim
    p^{\frac15}(\log p)^{\frac25}n^{-\frac25} \to 0\] with
  probability larger than \(1 - O(n^2p^{-9})\).
\end{corollary}

Finally, we would like to point out that the assumed spiked structure
on covariance plays an important role only in the theoretical
derivation. The proposed method still provides decent results in
simulation experiments when the spiked condition does not hold. %

\subsection{Extension to Nonlinear Classification Problems}
\label{sec:extens-nonl-class}
To address nonlinear classification challenges, a logical progression
would be to extend the proposed dynamic LDA methodology to the dynamic
QDA framework. Given that QDA provides greater flexibility in modeling
covariance matrices, it often facilitates improved separation between
two classes when the optimal discriminant boundary is nonlinear.

Specifically, we relax the equal covariance condition and assume
$\bX|{(Y=c, U=u)} \sim \mathcal{N}(\bmu^{(c)}(u),\bSigma^{(c)}(u))$,
$c=1,2$. In this scenario, the natural analogue for the common
covariance matrix is the pooled covariance of the two classes, so we
let \(\bSigma(u) : = \pi_1\bSigma^{(1)}(u) + \pi_2 \bSigma^{(2)}(u)\)
and use it to define the total covariance in~\eqref{eq:3}. The
rotation matrix $\bR(u)$ is defined the same way as in
Section~\ref{sec:dynam-discr-analys}. The following theorem shows that
under spiked assumptions on \(\bSigma^{(1)}(u)\) and
\(\bSigma^{(2)}(u)\), we can conduct dimension reduction without
information loss if an appropriate number of top principal components
are used for projection. Let $\{\lambda_j^{(c)}(u)\}_{j=1}^{p}$ denote
the eigenvalues of $\bSigma^{(c)}(u)$ in class $c$.

\begin{theorem}
  \label{thm:3}
  Assume a dynamic QDA model
  $\bX|{(Y=c, U=u)} \sim \mathcal{N}(\bmu^{(c)}(u),\bSigma^{(c)}(u))$,
  $c=1,2$, where the eigenvalues of $\bSigma^{(c)}(u)$ satisfy
  $\lambda_{1}^{(c)}(u)\geq\cdots\geq\lambda_{k_{c}}^{(c)}(u)>\lambda_{k_{c}+1}^{(c)}(u)=\cdots=\lambda_{p}^{(c)}(u)$,
  $c=1,2$, for some integers $k_{1},k_{2}<p$, and
  $\lambda_{p}^{(1)}(u)=\lambda_{p}^{(2)}(u)$. The optimal QDA rule is
  formulated by the first $k_{1}+k_{2}+1$ coordinates after linear
  transformation $\tilde\bx=\bR^{\top}(u)\bx$.
\end{theorem}

The implementation is straightforward. In Algorithm~\ref{alg:1}, we
use the same empirical total covariance for dimension reduction and
replace the LDA method with QDA for classification in Step 3. The
computation cost is only about 10\% higher than the LDA-based DSPCA in
our numerical experiments.

\section{Simulation Studies}
\label{sec4}

In this section, we conduct several simulation experiments to examine
the performance of our proposed method. To differentiate the variants
in Section~\ref{sec:implementation} and
Section~\ref{sec:extens-nonl-class}, we call them DSPCALDA and
DSPCAQDA respectively. The dynamic classification algorithm
DLPD~\citep{jiang2020dynamic} and its static counterpart Linear
Programming Discriminant \citep[LPD;][]{cai2011direct} are both
included for comparison. Principal Optimal Transport Direction
\citep[POTD;][]{meng2020sufficient} is a powerful supervised dimension
reduction tool that integrates optimal transport methods into the
sufficient dimension reduction framework, offering an appealing
alternative to the SPCA approach. Using the optimal subspace from
POTD, we construct a classifier that applies the standard LDA rule to
the projected data as an additional competitor in our numerical
analysis. Two other widely-used classifiers, Support Vector Machine
(SVM) with a linear kernel and K-Nearest Neighbors (KNN), are also
included. To conduct LPD, DLPD and POTD, we run the R code provided by
the authors under the recommended configurations. SVM and KNN are
performed using their implementations in the R packages \texttt{e1071}
and \texttt{class} respectively. For the static methods (LPD, POTD,
SVM, KNN), the index variable is not a valid input and is therefore
treated as an additional covariate. We also include an Oracle method,
which uses the optimal classification boundary calculated by the true
model parameters, as a benchmark.

The training and test data are generated as follows:
\begin{itemize}
\item Step 1: Generate dynamic indices independently from the standard
  uniform distribution
  $u_1, \dots, u_{n_1+n_2} \sim {\mathcal{U}}[0, 1]$.
\item Step 2: Sample
  $\bx_{i} \sim \mathcal{N}(\bmu^{(1)}(u_i),\bSigma^{(1)}(u_i))$ and
  assign $y_i=1$, for $i=1,\dots, n_1$.
\item Step 3: Sample
  $\bx_i\sim \mathcal{N}(\bmu^{(2)}(u_i),\bSigma^{(2)}(u_i))$ and
  assign $y_i=2$, for $i=n_1+1,\dots, n_1+n_2$.
\end{itemize}
We consider several settings for model parameters
$\bmu^{(c)}(u) = (\mu_{1}^{(c)}(u), \dots, \mu_{p}^{(c)}(u))^{\top}$
and $\bSigma^{(c)}(u)$. We include three models
from~\cite{jiang2020dynamic} (Models 1--3) and three more models
(Models 4--6) to have a closer inspection of different index variable
structures. In particular, Model 6 is designed with heteroscedasticity
so that the optimal classification rule is nonlinear.
\begin{itemize}
\item Model 1:
  $\mu_{1}^{(1)}(u)=\cdots=\mu_{p}^{(1)}(u)=1,
  \mu_{1}^{(2)}(u)=\cdots=\mu_{20}^{(2)}(u)=0,
  \mu_{21}^{(2)}(u)=\cdots=\mu_{p}^{(2)}(u)=1$ and
  $\bSigma^{(1)}(u)=\bSigma^{(2)}(u)=(0.5^{|i-j|})_{1\leq i,j \leq
    p}$.
\item Model 2:
  $\mu_{1}^{(1)}(u)=\cdots=\mu_{p}^{(1)}(u)=\exp(u),
  \mu_{1}^{(2)}(u)=\cdots=\mu_{20}^{(2)}(u)=u,
  \mu_{21}^{(2)}(u)=\cdots=\mu_{p}^{(2)}(u)=\exp(u)$ and
  $\bSigma^{(1)}(u)=\bSigma^{(2)}(u)=(u^{|i-j|})_{1\leq i,j \leq p}$.
\item Model 3:
  $\mu_{1}^{(1)}(u)=\cdots=\mu_{p}^{(1)}(u)=u,
  \mu_{1}^{(2)}(u)=\cdots=\mu_{20}^{(2)}(u)=-u,
  \mu_{21}^{(2)}(u)=\cdots=\mu_{p}^{(2)}(u)=u$ and
  $\bSigma^{(1)}(u)=\bSigma^{(2)}(u)=u\boldsymbol{1}_p\boldsymbol{1}_p^{\top}+(1-u)\bI_{p}$.
\item Model 4:
  $\mu_{1}^{(1)}(u)=\cdots=\mu_{p}^{(1)}(u)=u,
  \mu^{(2)}_{1}(u)=\cdots=\mu^{(2)}_{p-20}(u)=-u,
  \mu^{(2)}_{p-19}(u)=\cdots=\mu^{(2)}_{p}(u)=u$ and
  $\bSigma^{(1)}(u)=\bSigma^{(2)}(u)=u\boldsymbol{1}_p\boldsymbol{1}_p^{\top}+(1-u)\bI_{p}$.
\item Model 5:
  $\mu^{(1)}_{1}(u)=\cdots=\mu^{(1)}_{p}(u)=u,
  \mu^{(2)}_{1}(u)=\cdots=\mu^{(2)}_{p}(u)=\sin(4u)$, and
  $\bSigma^{(1)}(u)=\bSigma^{(2)}(u)=u\boldsymbol{1}_p\boldsymbol{1}_p^{\top}+(1-u)\bI_{p}$.
\item Model 6:
  $\mu^{(1)}_{1}(u)=\cdots=\mu^{(1)}_{p}(u)=u,
  \mu^{(2)}_{1}(u)=\cdots=\mu^{(2)}_{p-20}(u)=-u,
  \mu^{(2)}_{p-19}(u)=\cdots=\mu^{(2)}_{p}(u)=u$ and
  $\bSigma^{(1)}(u)=(u^{|i-j|})_{1\leq i,j \leq
    p},\bSigma^{(2)}(u)=u\boldsymbol{1}_p\boldsymbol{1}_p^{\top}+(1-u)\bI_{p}$.
\end{itemize}

Both the training and test data are set to include 100 observations in
each class, i.e., \(n_1 = n_2 = 100\). The dimension $p$ is chosen
from \(\{100, 150, 200\}\). For each scenario, we apply all methods to
100 independent replicates and report the means and standard errors of
misclassification rates in Tables~\ref{tab:1}--\ref{tab:6}. Under
Models 1 and 2, the normal vector
\(\bbeta(u) = \bSigma(u)^{-1}\bdelta(u)\) is sparse with
\(\|\bbeta(u)\|_0 = 21\), and the covariance structures are
considerably distant from a spiked model. Even though DSPCALDA and
DSPCAQDA are base on the projection of data to non-sparse directions,
their performance still slightly exceeds that of the
sparsity-promoting methods, i.e., LPD and DLPD\@. For Models 3, 4, and
5, where the sparsity assumption on $\bbeta(u)$ does not hold, LPD and
DLPD exhibit suboptimal performance and, in several instances, are
outperformed by SVM\@. In contrast, our proposed methods outperform
the competitors with misclassification rates very close to those of
the Oracle method. In Model 6, where the assumption of equal
covariance matrices does not hold, DSPCAQDA's misclassification rates
are significantly lower than those of the competing methods as
anticipated. Of all the static methods, POTD performs the best,
demonstrating competitive classification accuracy overall. However, it
is consistently outperformed by our methods in dynamic setups. These
simulation results lead us to conclude that DSPCA ensures robust
dynamic classification and consistently delivers the highest
prediction accuracy among all of the methods evaluated. Furthermore,
as demonstrated in Table~\ref{tab:7}, DSPCA achieves high
computational efficiency, comparable to that of static methods.

\begin{table}
  \captionsetup{font=small}
  \centering
  \scriptsize
  \caption{Average misclassification rates with standard errors for Model 1.}
  \label{tab:1}
  \begin{tabular}{ccccccccc}
    \(p\) & Oracle & POTD & SVM & KNN & LPD & DLPD & DSPCALDA & DSPCAQDA\\
    \midrule
    100 & 0.084(0.002) & \textbf{0.099}(0.002) & 0.155(0.003) & 0.176(0.003) & 0.110(0.002) & 0.111(0.002) & 0.100(0.002) & 0.100(0.002)\\
    150 & 0.080(0.002) & \textbf{0.100}(0.002) & 0.159(0.003) & 0.199(0.004) & 0.109(0.002) & 0.109(0.002) & 0.101(0.002) & 0.101(0.002)\\
    200 & 0.085(0.002) & 0.107(0.002) & 0.159(0.003) & 0.222(0.004) & 0.110(0.002) & 0.108(0.002) & \textbf{0.105}(0.002) & \textbf{0.105}(0.002)\\
  \end{tabular}
\end{table}

\begin{table}
  \captionsetup{font=small}
  \centering
  \scriptsize
  \caption{Average misclassification rates with standard errors for Model 2.}
  \label{tab:2}
  \begin{tabular}{ccccccccc}
    \(p\) & Oracle & POTD & SVM & KNN & LPD & DLPD & DSPCALDA & DSPCAQDA\\
    \midrule
    100 & 0.043(0.001) & 0.099(0.003) & 0.164(0.004) & 0.171(0.005) & 0.116(0.003) & 0.101(0.002) & \textbf{0.094}(0.002) & 0.096(0.002)\\
    150 & 0.041(0.001) & 0.096(0.003) & 0.160(0.004) & 0.187(0.006) & 0.115(0.002) & 0.105(0.002) & \textbf{0.094}(0.002) & 0.096(0.002)\\
    200 & 0.046(0.001) & \textbf{0.098}(0.002) & 0.156(0.003) & 0.220(0.006) & 0.118(0.002) & 0.113(0.002) & 0.104(0.002) & 0.107(0.002)\\
  \end{tabular}
\end{table}

\begin{table}
  \captionsetup{font=small}
  \centering
  \scriptsize
  \caption{Average misclassification rates with standard errors for Model 3.}
  \label{tab:3}
  \begin{tabular}{ccccccccc}
    \(p\) & Oracle & POTD & SVM & KNN & LPD & DLPD & DSPCALDA & DSPCAQDA\\
    \midrule
    100 & 0.085(0.002) & 0.122(0.003) & 0.146(0.003) & 0.213(0.003) & 0.144(0.003) & 0.138(0.003) & \textbf{0.104}(0.002) & 0.106(0.002)\\
    150 & 0.081(0.002) & 0.121(0.002) & 0.143(0.002) & 0.223(0.003) & 0.151(0.003) & 0.145(0.003) & \textbf{0.105}(0.002) & 0.108(0.002)\\
    200 & 0.092(0.002) & 0.119(0.002) & 0.139(0.003) & 0.220(0.004) & 0.149(0.002) & 0.145(0.003) & \textbf{0.110}(0.003) & 0.115(0.002)\\
  \end{tabular}
\end{table}

\begin{table}
  \captionsetup{font=small}
  \centering
  \scriptsize
  \caption{Average misclassification rates with standard errors for Model 4.}
  \label{tab:4}
  \begin{tabular}{ccccccccc}
    \(p\) & Oracle & POTD & SVM & KNN & LPD & DLPD & DSPCALDA & DSPCAQDA\\
    \midrule
    100 & 0.079(0.002) & 0.119(0.002) & 0.136(0.003) & 0.198(0.003) & 0.141(0.003) & 0.136(0.003) & \textbf{0.100}(0.002) & 0.104(0.002)\\
    150 & 0.079(0.002) & 0.119(0.002) & 0.128(0.002) & 0.199(0.003) & 0.147(0.002) & 0.144(0.003) & \textbf{0.103}(0.002) & 0.107(0.002)\\
    200 & 0.087(0.002) & 0.122(0.002) & 0.127(0.002) & 0.200(0.004) & 0.154(0.003) & 0.145(0.003) & \textbf{0.106}(0.002) & 0.111(0.003)\\
  \end{tabular}
\end{table}

\begin{table}
  \captionsetup{font=small}
  \centering
  \scriptsize
  \caption{Average misclassification rates with standard errors for Model 5.}
  \label{tab:5}
  \begin{tabular}{ccccccccc}
    \(p\) & Oracle & POTD & SVM & KNN & LPD & DLPD & DSPCALDA & DSPCAQDA\\
    \midrule
    100 & 0.332(0.003) & 0.485(0.004) & 0.485(0.004) & 0.490(0.003) & 0.480(0.004) & 0.373(0.004) & \textbf{0.346}(0.004) & 0.352(0.004)\\
    150 & 0.338(0.003) & 0.485(0.004) & 0.477(0.003) & 0.498(0.004) & 0.483(0.003) & 0.378(0.004) & \textbf{0.353}(0.004) & 0.354(0.004)\\
    200 & 0.340(0.003) & 0.493(0.004) & 0.482(0.003) & 0.500(0.004) & 0.489(0.003) & 0.376(0.004) & \textbf{0.352}(0.004) & 0.353(0.004)\\
  \end{tabular}
\end{table}

\begin{table}
  \captionsetup{font=small}
  \centering
  \scriptsize
  \caption{Average misclassification rates with standard errors for Model 6.}
  \label{tab:6}
  \begin{tabular}{ccccccccc}
    \(p\) & Oracle & POTD & SVM & KNN & LPD & DLPD & DSPCALDA & DSPCAQDA\\
    \midrule
    100 & 0.039(0.001) & 0.154(0.003) & 0.176(0.003) & 0.178(0.004) & 0.183(0.003) & 0.171(0.003) & 0.129(0.002) & \textbf{0.104}(0.002)\\
    150 & 0.036(0.001) & 0.147(0.002) & 0.167(0.003) & 0.253(0.005) & 0.175(0.003) & 0.168(0.003) & 0.124(0.003) & \textbf{0.101}(0.002)\\
    200 & 0.039(0.001) & 0.148(0.003) & 0.163(0.003) & 0.323(0.007) & 0.176(0.003) & 0.168(0.003) & 0.127(0.003) & \textbf{0.110}(0.003)\\
  \end{tabular}
\end{table}

\begin{table}
  \captionsetup{font=small}
  \centering
  \footnotesize
  \caption{Average computation time (in minutes) with standard errors for Model 6.}
  \label{tab:7}
  \begin{tabular}{cccccccc}
    \(p\) & POTD & SVM & KNN & LPD & DLPD & DSPCALDA & DSPCAQDA\\
    \midrule
    100 & 0.01(0.00) & 0.00(0.00) & 0.15(0.00) & 0.17(0.00) & 4.93(0.04) & 0.26(0.00) & 0.28(0.00)\\
    150 & 0.02(0.00) & 0.00(0.00) & 0.21(0.00) & 0.36(0.00) & 10.94(0.06) & 0.50(0.00) & 0.53(0.00)\\
    200 & 0.03(0.00) & 0.00(0.00) & 0.27(0.00) & 0.65(0.00) & 23.22(0.16) & 0.70(0.00) & 0.74(0.00)\\
  \end{tabular}
\end{table}

\section{Real Data Examples}
\label{sec5}
In this section, we evaluate the efficacy of our methodology through
two real data examples. For details regarding the methods to be
compared and their implementations, please refer to
Section~\ref{sec4}. Besides, we include a closely related static
method SPCALDA \citep{niu2015new}, which is conducted using the R
package of the same name under its default settings.

Breast cancer remains one of the most commonly diagnosed invasive
cancers among women worldwide. Our objective is to predict the
likelihood of its recurrence post-treatment within a specific time
frame. This study employs a binary classification approach akin to
that described by~\cite{wu2016regularized}. The first category
includes patients who have experienced metastases, relapse, or a
disease event within five years, while the second encompasses
individuals who have not encountered such events for a minimum of
seven years. To assess the effectiveness of our DSPCA methodology, we
compiled datasets GSE11121 and GSE1456 from the Gene Expression
Omnibus (GEO) database \citep{edgar2002gene}. Table~\ref{tab:8}
presents the basic information for these datasets.

\begin{table}
  \caption{Information for two breast cancer datasets.}
  \label{tab:8}
  \begin{minipage}{\textwidth}
    \centering
    \begin{tabular}{ccccc}
      Dataset & \makecell{Number \\ of genes} & \makecell{Number \\ of patients} & Class\footnote{t.dmgs represent the time for distant metastasis-free survival and e.dmfs is the corresponding event indicator.}    & \makecell{Number of patients \\ in each class}     \\
      \midrule
      GSE11121    & 22283  &125    &t.dmfs$\le$5y, e.dmfs=True           &28            \\
              &  &     &t.dmfs$>$7y, e.dmfs=False     & 97 \\
      GSE1456    &22283 & 105   &t.dmfs$\le$5y, relapse=True    & 33              \\
              & &  &t.dmfs$>7$y, relapse=False  & 72  \\
      \\

    \end{tabular}
  \end{minipage}
\end{table}

We partition each dataset into a training set and a test set, with 10
percent of the patients randomly selected for the latter based on
class proportions. Utilizing the binary response variable, two-sample
$t$-tests are employed to identify the top
$p\in\{50,\,100,\, 150,\, 200,\,250\}$ genes from the training set as
features. Note that different from the approach
in~\cite{jiang2020dynamic}, the data is partitioned prior to screening
to ensure exclusive reliance on information from the training set. For
the GSE11121 dataset, we follow the suggestion
in~\cite{jiang2020dynamic} and choose the tumor size as the index
variable. Furthermore, in alignment with findings from studies
\citep{rakha2010breast}, which highlight the prognostic significance
of tumor grade for breast cancer patients, the Elston tumor grade is
designated as the index variable for GSE1456. This ordered categorical
variable ranges from 1 to 3, with increasing grade numbers indicative
of faster-growing cancers and a higher propensity for spread. We
randomly split each dataset 100 times. The average misclassification
rates and standard errors on the test data, across 100 replications,
are documented in Tables~\ref{tab:9}--\ref{tab:10}.

Our results show that in general, dynamic methods, especially DSPCA,
perform better than the static ones although DLPD and LPD perform
similarly in many scenarios. This implies that the index variable
provides useful information for classification. The superiority of
prediction accuracy is more pronounced for DSPCAQDA in the case of
GSE1456, which indicates the advantage of nonlinear classification
techniques for complex data. The versatility of the DSPCA framework
allows users to choose from LDA and QDA after dimension reduction
depending on the structure of the dataset to achieve optimal
classification results. In summary, the DSPCA approach offers helpful
classification tools in analyzing modern complex data.

\begin{table}
  \captionsetup{font=small}
  \centering
  \scriptsize
  \caption{Average misclassification rates with standard errors for
    GSE11121.}
  \label{tab:9}
  \begin{tabular}{ccccccccc}
    \(p\) & POTD & SVM & KNN & LPD & SPCALDA & DLPD & DSPCALDA & DSPCAQDA\\
    \midrule
    50 & 0.234(0.009) & 0.237(0.011) & 0.246(0.011) & 0.243(0.010) & 0.228(0.009) & 0.258(0.011) & 0.214(0.009) & \textbf{0.194}(0.009)\\
    100 & 0.213(0.010) & 0.211(0.010) & 0.223(0.010) & 0.220(0.008) & 0.205(0.009) & 0.212(0.010) & 0.198(0.009) & \textbf{0.193}(0.009)\\
    150 & 0.204(0.008) & 0.199(0.008) & 0.211(0.010) & 0.224(0.009) & 0.199(0.009) & 0.214(0.011) & 0.201(0.008) & \textbf{0.182}(0.009)\\
    200 & 0.205(0.010) & 0.193(0.009) & 0.219(0.010) & 0.210(0.009) & 0.199(0.010) & 0.205(0.009) & \textbf{0.188}(0.008) & 0.189(0.010)\\
    250 & 0.200(0.009) & \textbf{0.189}(0.008) & 0.217(0.010) & 0.209(0.010) & 0.195(0.009) & 0.208(0.010) & 0.195(0.009) & 0.192(0.009)\\
  \end{tabular}
\end{table}

\begin{table}
  \captionsetup{font=small}
  \centering
  \scriptsize
  \caption{Average misclassification rates with standard errors for
    GSE1456.}
  \label{tab:10}
  \begin{tabular}{ccccccccc}
    \(p\) & POTD & SVM & KNN & LPD & SPCALDA & DLPD & DSPCALDA & DSPCAQDA\\
    \midrule
    50 & 0.322(0.013) & 0.346(0.014) & 0.305(0.012) & 0.321(0.012) & 0.295(0.012) & 0.310(0.012) & 0.285(0.012) & \textbf{0.267}(0.013)\\
    100 & 0.297(0.011) & 0.324(0.015) & 0.324(0.013) & 0.291(0.012) & 0.289(0.012) & 0.295(0.012) & 0.286(0.012) & \textbf{0.253}(0.013)\\
    150 & 0.288(0.011) & 0.334(0.013) & 0.338(0.013) & 0.285(0.012) & 0.297(0.012) & 0.302(0.013) & 0.281(0.011) & \textbf{0.271}(0.013)\\
    200 & 0.289(0.012) & 0.342(0.013) & 0.345(0.014) & 0.287(0.012) & 0.293(0.010) & 0.285(0.012) & 0.279(0.010) & \textbf{0.272}(0.013)\\
    250 & 0.290(0.012) & 0.336(0.013) & 0.348(0.014) & 0.299(0.012) & 0.313(0.011) & 0.315(0.012) & 0.291(0.012) & \textbf{0.275}(0.013)\\
  \end{tabular}
\end{table}

\section{Conclusion}
\label{sec:conc}
In this work, we introduce the DSPCA framework for high dimensional
classification, which offers new and more flexible tools for
non-static linear and quadratic discriminant analysis. The proposed
methods achieve high accuracy in classification by conducting
supervised dimension reduction in a dynamic fashion. Different from
existing dynamic classification techniques, our methods do not rely on
the sparsity conditions on the normal vectors of the optimal decision
boundaries. Our numerical studies show that the DSPCA-based methods
perform robust dynamic classification with high prediction accuracy
and computational efficiency, making it a competitive tool for
high-dimensional classification. An R package \texttt{DSPCA}
implementing our algorithm is available on GitHub~\citep{DSPCA}.

\section*{Appendix}

In this Appendix, we provide the definition of the quantity
\(B_n > 0\) introduced in Theorem~\ref{thm:2}. It depends on
\(n, p, h\) as well as other model parameters including
\(k, r, \Delta_1, \gamma, M\). We define \(B_n\) through the following
decomposition into \(B_{\RN{1}}\) and \(B_{\RN{2}}\):
\[B_n \coloneq B_{\RN{1}}M + B_{\RN{1}}^2 + B_{\RN{2}}.\]

\(B_{\RN{1}}\) originates from the estimation error of the conditional
first moments of \(\bX\), i.e.,
\(\bmu^{(c)}(u)= \E[\bX|Y = c, U = u]\). It is defined as
\begin{align*}
  B_{\RN{1}} &= \Delta_1\sqrt{\frac{r\log p}{hn}} + \Delta_1\frac{\sqrt{k}(\log p)^{3/2}}{hn} + h^2\Delta_1\sqrt{k\log p}\\
             &\quad + \gamma\sqrt{\frac{p\log p}{hn}} + \gamma\frac{\sqrt{p}(\log p)^{3/2}}{hn} + h^2\gamma\sqrt{p\log p}\\
             &\quad + M\sqrt{\frac{\log p}{hn}} + M\frac{\log p}{hn} + h^2M\sqrt{p}.
\end{align*}

\(B_{\RN{2}}\) originates from the estimation error of the conditional
second moments of \(\bX\), i.e.,
\(\bPi^{(c)}(u) = \E[\bX\bX^{\top}|Y = c, U = u] = \bSigma(u) +
\bmu^{(c)}(u)\bmu^{(c)\top}(u)\). It is defined as
\begin{align*}
  B_{\RN{2}} &= \Delta_1^2\sqrt{\frac{k\log p}{hn}} + \Delta_1^2\frac{k(\log p)^2}{hn} + h^2\Delta_1^2kp\\
             &\quad + \gamma^2\sqrt{\frac{p\log p}{hn}} + \gamma^2\frac{p(\log p)^2}{hn} + h^2\gamma^2 p \log p\\
             &\quad + M^2\sqrt{\frac{\log p}{hn}} + M^2\frac{\log p}{hn} + h^2M^2p\\
             &\quad + \Delta_1\gamma\sqrt{\frac{p\log p}{hn}} + \Delta_1M\sqrt{\frac{k\log p}{hn}} + \gamma M\sqrt{\frac{p\log p}{hn}}.
\end{align*}

\section*{Supplementary Material}
\label{sec:suppl-mater}
The Supplementary Material consists of auxiliary Lemmas S.1--S.5 and
the proofs of the lemmas, theorems, and corollaries.

\section*{Acknowledgment}
\label{sec:acknowledgment}
The authors thank the Editor, Associate Editor, and three referees for
their helpful comments. Wu was supported by the National Institutes of
Health grant 1R21AG074205-01. Hao was supported by the National
Science Foundation grant DMS-2245381 and the Simons Foundation grant
524432. Zhang was supported by the National Science Foundation grant
DMR 2242925 and the National Institutes of Health grant 1R01 CA260399.
Additional support was provided by NYU IT High Performance Computing
resources, services, and staff expertise.

\section*{Disclosure Statement}
\label{sec:disclosure-statement}
The authors report there are no competing interests to declare.

\putbib
\end{bibunit}

\newpage
\begin{center}
  {\LARGE\bfseries Supplementary Material}
\end{center}

\vspace{1cm}

\setcounter{lemma}{0}
\setcounter{equation}{0}
\renewcommand{\thelemma}{S.\arabic{lemma}}
\renewcommand{\theequation}{S.\arabic{equation}}

\begin{bibunit}
We illustrate the proof of the theoretical results,
Theorem~\ref{thm:2}, Corollary~\ref{cor:1} and Theorem~\ref{thm:3} in
this Supplementary Material. Theorem~\ref{thm:1} is proved in a
similar way to Theorem~\ref{thm:3}.

First, we restate the truncated matrix Bernstein
inequality~\citep{chen2021spectral} in the vector form and symmetric
matrix form, which are more convenient for our application.

\begin{lemma}
  \label{lem:s1}
  Let \(\{\bN_i\}_{1\leq i\leq n}\) be a sequence of independent and
  identically distributed (iid) length-\(p\) random vectors. Suppose
  that for all \(1 \leq i \leq n\),
  \[\rP\{\|\bN_i - \E \bN_i\| \geq L\} \leq q_0,\]
  \[\left\|\E[\bN_i 1_{\|\bN_i\| \geq L}]\right\| \leq q_1\] hold
  for \(0 \leq q_0, q_1 \leq 1\). In addition, define the matrix
  variance statistic \(V\) as
  \[V \coloneq n\tr(\E[(\bN_i - \E\bN_i)(\bN_i - \E\bN_i)^{\top}]).\]
  Then for \(a \geq 2\), with probability exceeding
  \(1 - 2p^{-a + 1} - nq_0\) it holds that
  \[\left\|\sum_{i = 1}^n(\bN_i - \E\bN_i)\right\| \leq \sqrt{2aV
      \log p} + \frac{2a}{3}L\log p + nq_1.\]
\end{lemma}
\begin{lemma}
  \label{lem:s2}
  Let \(\{\bM_i\}_{1\leq i\leq n}\) be a sequence of iid symmetric
  \(p \times p\) random matrices. Suppose that for all
  \(1 \leq i \leq n\),
  \[\rP\{\|\bM_i - \E\bM_i\| \geq L\} \leq q_0,\]
  \[\left\|\E[\bM_i 1_{\|\bM_i\| \geq L}]\right\| \leq q_1\] hold
  for \(0 \leq q_0, q_1 \leq 1\). In addition, define the matrix
  variance statistic \(V\) as
  \[V \coloneq n\left\|\E[(\bM_i - \E\bM_i)^2]\right\|.\] Then for
  \(a \geq 2\), with probability exceeding
  \(1 - 2p^{-a + 1} - nq_0\) it holds that
  \[\left\|\sum_{i = 1}^n(\bM_i - \E\bM_i)\right\| \leq \sqrt{2aV
      \log p} + \frac{2a}{3}L\log p + nq_1.\]
\end{lemma}

In the sequel, for any function depending on the variable \(u\), we
will omit \(u\) whenever no ambiguity shall arise. For example, we
will write \(\bSigma\) for \(\bSigma(u)\). We will also abuse the
notation \(\rP(a_n \gtrsim b_n) \lesssim \cdots\) to mean that there
exists some \(C_1, C_2 > 0\) such that
\(\rP(|a_n| \geq C_1 |b_n|) \leq C_2\cdots\) for all sufficiently
large \(n\).

\begin{lemma}
  \label{lem:s3}
  Suppose that \(\frac{\log p}{hn} \to 0\), \(h \to 0\),
  \(\rho = O(1)\). Under Assumptions~\ref{asp:1}--\ref{asp:6}, it
  holds that
  \[\rP\left(\sup_{u \in [0, 1]}\|\hat{\bSigma}_{\rho}^{tot}(u) -
      \bSigma_{\rho}^{tot}(u)\| \gtrsim B_{\RN{1}}M + B_{\RN{1}}^2 +
      B_{\RN{2}}\right) \lesssim h^{-4}np^{-11.5},\] where
  \begin{align*}
    B_{\RN{1}} &= \Delta_1\sqrt{\frac{r\log p}{hn}} + \Delta_1\frac{\sqrt{k}(\log p)^{3/2}}{hn} + h^2\Delta_1\sqrt{k\log p}\\
               &\quad + \gamma\sqrt{\frac{p\log p}{hn}} + \gamma\frac{\sqrt{p}(\log p)^{3/2}}{hn} + h^2\gamma\sqrt{p\log p}\\
               &\quad + M\sqrt{\frac{\log p}{hn}} + M\frac{\log p}{hn} + h^2M\sqrt{p}, 
  \end{align*}
  and
  \begin{align*}
    B_{\RN{2}} &= \Delta_1^2\sqrt{\frac{k\log p}{hn}} + \Delta_1^2\frac{k(\log p)^2}{hn} + h^2\Delta_1^2kp\\
               &\quad + \gamma^2\sqrt{\frac{p\log p}{hn}} + \gamma^2\frac{p(\log p)^2}{hn} + h^2\gamma^2 p \log p\\
               &\quad + M^2\sqrt{\frac{\log p}{hn}} + M^2\frac{\log p}{hn} + h^2M^2p\\
               &\quad + \Delta_1\gamma\sqrt{\frac{p\log p}{hn}} + \Delta_1M\sqrt{\frac{k\log p}{hn}} + \gamma M\sqrt{\frac{p\log p}{hn}}.
  \end{align*}
\end{lemma}
\begin{myproof}
  We first introduce some notations. Let
  \(w_i(u) = K_h(u_i - u)/n_1\),
  \(\bPi^{(1)} = \bSigma + \bmu^{(1)}\bmu^{(1)\top}\),
  \(\bPi^{(2)} = \bSigma + \bmu^{(2)}\bmu^{(2)\top}\),
  \(\hat\bPi^{(1)} = \frac{\sum_{i:y_i = 1}w_i\bx_i
    \bx_i^{\top}}{\sum_{i:y_i = 1}w_i}\), and
  \(\hat\bPi^{(2)} = \frac{\sum_{i:y_i = 2}w_i\bx_i
    \bx_i^{\top}}{\sum_{i:y_i = 2}w_i}\). Then
  \[\bSigma^{tot}_{\rho} = (\pi_1\bPi^{(1)} + \pi_2\bPi^{(2)}) -
    \left(\pi_1\bmu^{(1)}\bmu^{(1)\top} +
      \pi_2\bmu^{(2)}\bmu^{(2)\top}\right) + \rho (\bmu^{(1)} -
    \bmu^{(2)})(\bmu^{(1)} - \bmu^{(2)})^{\top}\]
  \begin{align*}
    \hat\bSigma^{tot}_{\rho} &= \frac{n_1}{n}\left(\hat\bPi^{(1)} - \hat\bmu^{(1)} \hat\bmu^{(1)\top}\right) + \frac{n_2}{n}\left(\hat\bPi^{(2)} - \hat\bmu^{(2)} \hat\bmu^{(2)\top}\right) + \rho (\hat\bmu^{(1)} - \hat\bmu^{(2)})(\hat\bmu^{(1)} - \hat\bmu^{(2)})^{\top}\\
                             &= \left(\frac{n_1}{n}\hat\bPi^{(1)} + \frac{n_2}{n}\hat\bPi^{(2)}\right) - \left(\frac{n_1}{n}\hat\bmu^{(1)} \hat\bmu^{(1)\top} + \frac{n_2}{n}\hat\bmu^{(2)} \hat\bmu^{(2)\top}\right) + \rho (\hat\bmu^{(1)} - \hat\bmu^{(2)})(\hat\bmu^{(1)} - \hat\bmu^{(2)})^{\top}
  \end{align*}

  We claim the following inequalities are true:
  \begin{equation}
    \label{eq:s1}
    \rP\left(\left|\frac{n_1}{n} - \pi_1\right| \gtrsim \sqrt{\frac{\log p}{n}}\right) \leq p^{-11.5},
  \end{equation}
  \begin{equation}
    \label{eq:s2}
    \rP\left(\sup_{u \in [0, 1]} \|\hat \bmu^{(1)} - \bmu^{(1)}\| \gtrsim B_{\RN{1}}\right)\lesssim h^{-4}np^{-11.5},
  \end{equation}
  \begin{equation}
    \label{eq:s3}
    \rP\left(\sup_{u \in [0, 1]} \|\hat\bPi^{(1)} - \bPi^{(1)}\| \gtrsim B_{\RN{2}}\right) \lesssim h^{-4}np^{-11.5},
  \end{equation}

  Inequality~\eqref{eq:s1} is a direct consequence of Hoeffding's
  inequality. Since~\eqref{eq:s2} and~\eqref{eq:s3} only concerns the
  probability distributions of \(\hat \bmu^{(1)}(u)\) and
  \(\hat \bPi^{(1)}(u)\), it does not matter how they are constructed
  as random variables. Next, we will construct a random sample
  \((\bx_i, u_i, y_i)_{i = 1}^n\) that has the desired joint
  distribution and work with \((\bx_i, u_i, y_i)_{i = 1}^n\)
  towards~\eqref{eq:s2} and~\eqref{eq:s3}, without loss of generality.

  Recall the following decomposition of \(\bSigma(u)\) as
  in~\eqref{eq:8}
  \[\bSigma(u) = \bL(u)\bL(u)^{\top} + \sigma(u)^2\bI_p.\]
  For \(1 \leq i \leq n\), let
  \[\bx_i = \bL(u_i) \btheta_i + \sigma(u_i)\bEta_i +
    \bmu^{(y_i)}(u_i),\] where \(u_i\), \(y_i\), \(\btheta_i\) and
  \(\bEta_i\) are generated independently according to
  \[u_i \overset{\mathrm{iid}}{\sim} f_U, \; y_i - 1
    \overset{\mathrm{iid}}{\sim} \ber(\pi_2), \; \btheta_i
    \overset{\mathrm{iid}}{\sim} \cN(\bzero, \bI_k), \; \bEta_i
    \overset{\mathrm{iid}}{\sim} \cN(\bzero, \bI_p).\] Let
  \(\by = (y_1, \dots, y_n)\) and \(\{1, 2\}^{(n)}\) denote the set of
  all length-\(n\) sequences taking values in \(\{1, 2\}\). It is easy
  to see that given \(\bc = (c_1, \dots, c_n) \in \{1, 2\}^{(n)}\),
  when conditioned on the event \(\by = \bc\), for \(c = 1, 2\),
  \(\{(\bx_i, u_i)\}_{i: c_i = c}\) are iid, with
  \begin{equation}
    \label{eq:s4}
    u_i \overset{\mathrm{iid}}{\sim} f_U, \; (\bx_i | u_i = u) \overset{\mathrm{iid}}{\sim} \cN(\bmu^{(c)}(u),
    \bSigma(u)).
  \end{equation}
  To simplify notation, we define the summation symbols
  \(\sum_{\bc = 1} \coloneq \sum_{i:c_i = 1}\) and the conditional
  probability symbol
  \(\rP_{\bc}(\cdot) \coloneq \rP(\cdot | \by = \bc)\). Furthermore,
  let \(n_{\bc} = \sum_{\bc = 1}1\) and
  \(w(u) = \sum_{\bc = 1}w_i(u)\). We also use the notations
  \(B_{\RN{1}}(n)\) and \(B_{\RN{2}}(n)\) to emphasize \(B_{\RN{1}}\)
  and \(B_{\RN{2}}\)'s dependency on \(n\).

  We claim (with the proofs deferred to improve readability) that
  \begin{equation}
    \label{eq:s5}
    \rP_{\bc}\left(\sup_{u \in [0, 1]} \left\|\sum_{\bc = 1}\frac{w_i}{w}\bx_i - \bmu^{(1)}\right\|\gtrsim B_{\RN{1}}(n_{\bc})\right) \lesssim h^{-4}n_{\bc}p^{-11.5},
  \end{equation}
  \begin{equation}
    \label{eq:s6}
    \rP_{\bc}\left(\sup_{u \in [0, 1]} \left\|\sum_{\bc = 1} \frac{w_i}{w}\bx_{i}\bx_{i}^{\top} - \bPi^{(1)}\right\| \gtrsim B_{\RN{2}}(n_{\bc}) \right) \lesssim h^{-4}n_{\bc}p^{-11.5},
  \end{equation}

  Conditioned on the event \(\{\by = \bc\}\) where
  \(\bc \in \{1, 2\}^{(n)}\), it is clear that \(n_1 = n_{\bc}\),
  \begin{equation}
    \label{eq:s7}
    \hat{\bmu}^{(1)} = \sum_{\bc = 1}\frac{w_i}{w}\bx_i,
  \end{equation}
  and
  \[\hat{\bPi}^{(1)} = \sum_{\bc = 1} \frac{w_i}{w}\bx_{i}\bx_{i}^{\top}.\]
  Thus
  \begin{align*}
    &\quad\rP\left(\sup_{u \in [0, 1]} \|\hat \bmu^{(1)} - \bmu^{(1)}\| \gtrsim B_{\RN{1}}(n)\right)\\
    &=\rP\left(\left\{\sup_{u \in [0, 1]} \|\hat \bmu^{(1)} - \bmu^{(1)}\| \gtrsim B_{\RN{1}}(n)\right\}\cap \left\{n_1 < \frac{\pi_1}{2}n\right\}\right)\\
    &\quad + \rP\left(\left\{\sup_{u \in [0, 1]} \|\hat \bmu^{(1)} - \bmu^{(1)}\| \gtrsim B_{\RN{1}}(n)\right\}\cap \left\{n_1 \geq \frac{\pi_1}{2}n\right\}\right)\\
    &\leq \rP\left(n_1 < \frac{\pi_1}{2}n\right)\\
    &\quad + \sum_{\bc \in \{1, 2\}^{(n)}}\rP_{\bc}\left(\left\{\sup_{u \in [0, 1]} \|\hat \bmu^{(1)} - \bmu^{(1)}\| \gtrsim B_{\RN{1}}(n)\right\} \cap \left\{n_{\bc} \geq \frac{\pi_1}{2}n\right\}\right)\rP(\by = \bc)\\
    &\overset{\mathrm{(i)}}{\leq} p^{-11.5} + \sum_{\bc \in \{1, 2\}^{(n)}:n_{\bc} \geq \frac{\pi_1}{2}n}\rP_{\bc}\left(\sup_{u \in [0, 1]} \left\|\sum_{\bc = 1}\frac{w_i}{w}\bx_i - \bmu^{(1)}\right\| \gtrsim B_{\RN{1}}(n)\right)\rP(\by = \bc)\\
    &\overset{\mathrm{(ii)}}{\lesssim} p^{-11.5} + h^{-4}n_{\bc}p^{-11.5}\sum_{\bc \in \{1, 2\}^{(n)}:n_{\bc} \geq \frac{\pi_1}{2}n} P(\by = \bc)\\
    &\lesssim h^{-4}np^{-11.5},
  \end{align*}
  where (i) follows from~\eqref{eq:s1} and~\eqref{eq:s7}, and (ii)
  follows from~\eqref{eq:s5}. Similarly,
  \begin{align*}
    &\quad \rP\left(\sup_{u \in [0, 1]} \|\hat\bPi^{(1)} - \bPi^{(1)}\| \gtrsim B_{\RN{2}}(n)\right)\\
    &\leq p^{-11.5} + \sum_{\bc \in \{1, 2\}^{(n)}:n_{\bc} \geq \frac{\pi_1}{2}n}\rP_{\bc}\left(\sup_{u \in [0, 1]} \left\|\sum_{\bc = 1} \frac{w_i}{w}\bx_{i}\bx_{i}^{\top} - \bPi^{(1)}\right\|\gtrsim B_{\RN{2}}(n)\right)\rP(\by = \bc)\\
    &\overset{\mathrm{(i)}}{\lesssim} p^{-11.5} + h^{-4}n_{\bc}p^{-11.5}\sum_{\bc \in \{1, 2\}^{(n)}:n_{\bc} \geq \frac{\pi_1}{2}n} P(\by = \bc)\\
    &\lesssim h^{-4}np^{-11.5},
  \end{align*}
  where (i) follows from~\eqref{eq:s6}.

  Now that we have proved~\eqref{eq:s2} and~\eqref{eq:s3}, we return
  to the main proof. Note that
  \begin{align*}
    &\quad\left\|\frac{n_1}{n}\hat\bmu^{(1)}\hat\bmu^{(1)\top} - \pi_1\bmu^{(1)}\bmu^{(1)\top}\right\| \leq \left\| \frac{n_1}{n}\left(\hat\bmu^{(1)}\hat\bmu^{(1)\top} - \bmu^{(1)}\bmu^{(1)\top}\right)\right\| + \left\|\left(\frac{n_1}{n} - \pi_1\right)\bmu^{(1)}\bmu^{(1)\top}\right\|\\
    &\leq \|\hat\bmu^{(1)}\hat\bmu^{(1)\top} - \bmu^{(1)}\bmu^{(1)\top}\| + M^2\left|\frac{n_1}{n} - \pi_1\right|\\
    &\leq \|\left(\hat\bmu^{(1)} - \bmu^{(1)}\right)\left(\hat\bmu^{(1)} - \bmu^{(1)}\right)^{\top}\| + \|\bmu^{(1)}\hat\bmu^{(1)\top} + \hat\bmu^{(1)}\bmu^{(1)\top} - 2\bmu^{(1)}\bmu^{(1)\top}\| + M^2\left|\frac{n_1}{n} - \pi_1\right|\\
    &\leq \|\hat\bmu^{(1)} - \bmu^{(1)}\|^2 + \|\bmu^{(1)}\left(\hat\bmu^{(1)} - \bmu^{(1)}\right)^{\top}\| + \|\left(\hat\bmu^{(1)} - \bmu^{(1)}\right)\bmu^{(1)\top}\| + M^2\left|\frac{n_1}{n} - \pi_1\right|\\
    &\lesssim M \|\hat\bmu^{(1)} - \bmu^{(1)}\| + \|\hat\bmu^{(1)} - \bmu^{(1)}\|^2 + M^2\left|\frac{n_1}{n} - \pi_1\right|
  \end{align*}
  and
  \begin{align*}
    &\quad\left\|\frac{n_1}{n}\hat\bPi^{(1)} - \pi_1\bPi^{(1)}\right\| \leq \left\|\frac{n_1}{n}(\hat\bPi^{(1)} - \bPi^{(1)})\right\| + \left\|\left(\frac{n_1}{n} - \pi_1\right)\bPi^{(1)}\right\|\\
    &\leq \|\hat\bPi^{(1)} - \bPi^{(1)}\| + \|\bL\bL^{\top} + \sigma^2\bI_p + \bmu^{(1)}\bmu^{(1)\top}\|\left|\frac{n_1}{n} - \pi_1\right|\\
    &\leq \|\hat\bPi^{(1)} - \bPi^{(1)}\| + (\Delta_1^2 + \gamma^2 + M^2)\left|\frac{n_1}{n} - \pi_1\right|.
  \end{align*}
  
  By~\eqref{eq:s1}--\eqref{eq:s3} and the union bound, it holds that
  \[\sup_{u \in [0,
      1]}\left\|\frac{n_1}{n}\hat\bmu^{(1)}\hat\bmu^{(1)\top} -
      \pi_1\bmu^{(1)}\bmu^{(1)\top}\right\| \lesssim B_{\RN{1}}M +
    B_{\RN{1}}^2\] and
  \[\sup_{u \in [0, 1]}\left\|\frac{n_1}{n}\hat\bPi^{(1)} -
      \pi_1\bPi^{(1)}\right\| \lesssim B_{\RN{2}}\] with probability
  exceeding \(1 - O(h^{-4}np^{-11.5})\).

  Furthermore, similar bounds can be derived for
  \(\|\frac{n_2}{n}\hat\bmu^{(2)}\hat\bmu^{(2)\top} -
  \pi_1\bmu^{(2)}\bmu^{(2)\top}\|\),
  \(\|\frac{n_2}{n}\hat\bPi^{(2)} - \pi_1\bPi^{(2)}\|\) and
  \(\|(\hat\bmu^{(1)} - \hat\bmu^{(2)})(\hat\bmu^{(1)} -
  \hat\bmu^{(2)})^{\top} - (\bmu^{(1)} - \bmu^{(2)})(\bmu^{(1)} -
  \bmu^{(2)})^{\top}\|\). Combining these bounds completes the proof.
\end{myproof}

To prove~\eqref{eq:s5} and~\eqref{eq:s6}, we first rephrase Lemma A.1
of~\cite{jiang2020dynamic} in a form that is easier to use for our
purposes.
\begin{lemma}
  \label{lem:s4}
  Under Assumptions~\ref{asp:1} and~\ref{asp:6}, it holds that
  \[\rP_{\bc}\left(\sup_{u \in [0, 1]}|w - f_U(u)| \gtrsim
      \sqrt{\frac{\log p}{hn_{\bc}}} + h^2\right) \lesssim
    h^{-4}p^{-11.5}.\]
\end{lemma}
To obtain the inequality in Lemma~\ref{lem:s4}, just set
\(\epsilon_n = \sqrt{11.5\log p/(C_2hn_{\bc})}\), \(d = 1\), and
replace the rate \((\log p/n)^{1/(4 + d)}\) with the original
parameter \(h\) for Lemma A.1 of~\cite{jiang2020dynamic}. When
\(\frac{\log p}{hn_{\bc}} \to 0\) and \(h \to 0\), Lemma~\ref{lem:s4}
shows that \(w\) is bounded away from 0 with high probability. This is
crucial because \(w\) occurs as the denominator in the Nadaraya-Watson
estimator.

Next, throughout the proof
of~\eqref{eq:s5},~\eqref{eq:s6},~\eqref{eq:s15} and~\eqref{eq:s16}, we
assume all statements involving probability and randomness are
conditioned on the event \(\{\by = \bc\}\) where
\(\bc \in \{1, 2\}^{(n)}\), e.g., \(\rP(\cdot)\) stands for
\(\rP(\cdot|\by = \bc)\), \(\E[\cdot]\) stands for
\(\E[\cdot|\by = \bc]\), \(\bmu^{(y_i)}(u_i) = \bmu^{(c_i)}(u_i)\),
\(n_1 = n_{\bc}\), etc.
\begin{myproof}[Proof of~\eqref{eq:s5}]
  Note that
  \[\sum_{\bc = 1} \frac{w_i}{w}\bx_i - \bmu^{(1)}=\sum_{\bc = 1}
    \frac{w_i}{w}\bL(u_i) \btheta_i + \sum_{\bc = 1}
    \frac{w_i}{w}\sigma(u_i)\bEta_i + \left(\sum_{\bc = 1}
      \frac{w_i}{w}\bmu^{(1)}(u_i) - \bmu^{(1)}(u)\right).\] It
  suffices to prove the following three inequalities:
  \begin{multline}
    \label{eq:s8}
    \rP\Bigg(\sup_{u \in [0, 1]}\left\|\sum_{\bc = 1}\frac{w_i}{w}\bL(u_i)\btheta_i\right\|\\
    \gtrsim \Delta_1\sqrt{\frac{r\log p}{hn_{\bc}}} + \Delta_1\frac{\sqrt{k}(\log p)^{3/2}}{hn_{\bc}} + h^2\Delta_1\sqrt{k\log p} \Bigg) \lesssim h^{-4}n_{\bc}p^{-11.5},
  \end{multline}
  \begin{multline}
    \label{eq:s9}
    \rP\Bigg(\sup_{u \in [0, 1]}\left\|\sum_{\bc = 1}\frac{w_i}{w}\sigma(u_i)\bEta_i\right\|\\
    \gtrsim \gamma\sqrt{\frac{p\log p}{hn_{\bc}}} + \gamma\frac{\sqrt{p}(\log p)^{3/2}}{hn_{\bc}} + h^2\gamma\sqrt{p\log p} \Bigg) \lesssim h^{-4}n_{\bc}p^{-11.5},
  \end{multline}
  \begin{multline}
    \label{eq:s10}
    \rP\Bigg(\sup_{u \in [0, 1]}\left\|\sum_{\bc = 1}\frac{w_i}{w}\bmu^{(1)}(u_i) - \bmu^{(1)}(u)\right\|\\
    \gtrsim M\sqrt{\frac{\log p}{hn_{\bc}}} + M\frac{\log p}{hn_{\bc}} + h^2M\sqrt{p}\Bigg) \lesssim h^{-4}p^{-11.5}
  \end{multline}
  Here we only provide a proof of~\eqref{eq:s8}, which can be easily
  adapted for~\eqref{eq:s9}. Inequality~\eqref{eq:s10} can be proved
  via a similar approach to the one used for the proof
  of~\eqref{eq:s14}.

  By Assumption~\ref{asp:1}, \(f_U \geq C_U > 0\) on \([0, 1]\). With
  Lemma~\ref{lem:s4} this implies
  \(\rP(\inf_{u \in [0, 1]}w \gtrsim C_U/2) \lesssim
  h^{-4}p^{-11.5}\). Thus it suffices to show
  \[\rP\left(\sup_{u \in [0, 1]}\left\|\sum_{\bc = 1} \bN_i\right\| \gtrsim
      \Delta_1\sqrt{\frac{r\log p}{hn_{\bc}}} +
      \Delta_1\frac{\sqrt{k}(\log p)^{3/2}}{hn_{\bc}} +
      h^2\Delta_1\sqrt{k\log p} \right) \lesssim
    h^{-4}n_{\bc}p^{-11.5},\] where
  \(\bN_i(u) = w_i(u)\bL(u_i)\btheta_i\).

  Recall that
  \(\bL(u) = \bQ_1(u)\diag(\sqrt{\lambda_1(u) - \sigma(u)^2}, \dots,
  \sqrt{\lambda_k(u) - \sigma(u)^2})\). In view of
  Assumption~\ref{asp:4}, \(\bQ_1(u) = \bW \tilde\bQ_1(u)\), where
  \(\bW\) is a \(p \times r\) matrix whose columns form an orthonormal
  basis of \(W\), and \(\tilde \bQ_1(u)\) is an \(r \times k\) matrix
  with orthonormal columns. Define
  \(\tilde\bL(u) = \tilde\bQ_1(u) \diag(\sqrt{\lambda_1(u) -
    \sigma(u)^2}, \dots, \sqrt{\lambda_k(u) - \sigma(u)^2})\), then
  \(\bL(u) = \bW \tilde\bL(u)\). Clearly \(\|\bW\| = 1\) and by
  Assumption~\ref{asp:3},
  \(\|\tilde\bL(u)\| = \|\bL(u)\| \leq \Delta_1\).

  Next, a two-step procedure is employed, where we first derive an
  upper bound of \(\|\sum_{\bc = 1}\bN_i(u)\|\) for each fixed \(u\)
  using Lemma~\ref{lem:s1}, and then extrapolate the result to all
  \(u \in [0, 1]\) using a grid-based argument.

  Step 1, assume \(u \in [0, 1]\) is fixed. First note that
  \(\E[\bN_i] = \bzero\). Since
  \(\bN_i = w_i\bW\tilde\bL(u_i)\btheta_i\) where
  \(w_i = K_h(u_i - u)/n_{\bc}\), we have
  \(\|\bN_i\| \leq |w_i|\|\bW\|\|\tilde\bL(u_i)\|\|\btheta_i\| \leq
  \frac{C_K\Delta_1}{n_{\bc}h}\|\btheta_i\| \leq \frac{C_K\Delta_1
    \sqrt{k}}{n_{\bc}h}\|\btheta_i\|_{\infty}\) by
  Assumption~\ref{asp:6}. By the normality of \(\btheta_i\), we can
  easily verify
  \(\rP(\|\btheta_i\|_{\infty} \leq 5\sqrt{\log p}) \geq 1 -
  p^{-11.5}\). As a result,
  \[\rP\left(\|\bN_i\| \geq \frac{5C_K\Delta_1 \sqrt{k}}{n_{\bc}h}
      \sqrt{\log p}\right) \leq p^{-11.5}.\] To apply
  Lemma~\ref{lem:s1}, we define
  \(L \coloneq \frac{5 C_K\Delta_1 \sqrt{k}}{n_{\bc}h} \sqrt{\log p}\)
  and \(q_0 \coloneq p^{-11.5}\).

  Additionally, the symmetric properties of the Gaussian distribution
  implies \(\E[\bN_i 1_{\|\bN_i\| \geq L}] = \bzero\). Thus
  \(q_1 \coloneq 0\).

  The matrix variance statistic \(V\) can be bounded as follows:
  \begin{align*}
    &\quad V = n_{\bc} \tr(\E[\bN_i\bN_i^{\top}])\\
    &= n_{\bc} \tr(\bW \E[w_i^2\tilde\bL(u_i)\btheta_i\btheta_i^{\top}\tilde\bL(u_i)^{\top}]\bW^{\top}) = n_{\bc} \tr(\bW^{\top}\bW \E[w_i^2\tilde\bL(u_i)\btheta_i\btheta_i^{\top}\tilde\bL(u_i)^{\top}])\\
    &= n_{\bc} \tr(\E[w_i^2\tilde\bL(u_i)\btheta_i\btheta_i^{\top}\tilde\bL(u_i)^{\top}]) \leq n_{\bc}r\|\E[w_i^2\tilde\bL(u_i)\btheta_i\btheta_i^{\top}\tilde\bL(u_i)^{\top}]\|\\
    &= n_{\bc} r \|\E\{\E[w_i^2\tilde\bL(u_i)\btheta_i\btheta_i^{\top}\tilde\bL(u_i)^{\top}|u_i]\}\| = n_{\bc}r\|\E[w_i^2\tilde\bL(u_i)\tilde\bL(u_i)^{\top}]\|\\
    &\leq n_{\bc}r \E[w_i^2\|\tilde\bL(u_i)\|^2] \leq n_{\bc} r \Delta_1^2 \E[w_i^2] \leq \frac{\tilde C_U C_K\Delta_1^2}{n_{\bc}h}r,
  \end{align*}
  where the last inequality is a consequence of Assumption~\ref{asp:1}
  and \ref{asp:6} shown as follows:
  \begin{equation}
    \label{eq:s11}
    \begin{aligned}
      \E[w_i^2] &= \int_{\R} \frac{1}{n_{\bc}^2h^2} K\left(\frac{v - u}{h}\right)^2 f_U(v) dv\\
                &= \frac{1}{n_{\bc}^2h} \int_{\R} K(\nu)^2 f_U(u + h\nu)d\nu\\
                & \leq \frac{\tilde C_U}{n_{\bc}^2h} \int_{\R} K(\nu)^2 d\nu \leq \frac{\tilde C_U C_K}{n_{\bc}^2h}.
    \end{aligned}
  \end{equation}
  Now we can apply Lemma~\ref{lem:s1} and conclude with probability
  exceeding \(1-O(n_{\bc}p^{-11.5})\)
  \begin{equation}
    \label{eq:s12}
    \left\|\sum_{\bc = 1}\bN_i(u)\right\| \lesssim \Delta_1\sqrt{\frac{r\log p}{hn_{\bc}}} + \Delta_1\frac{\sqrt{k}(\log p)^{3/2}}{hn_{\bc}}.
  \end{equation}

  Step 2, we construct a equally spaced grid in \([0, 1]\) with
  \(\lceil 1/h^4\rceil + 1\) grid points
  \(\{v_l: 0 \leq l \leq \lceil 1/h^4 \rceil\}\) (including end
  points). The distance between neighboring grid points is thus less
  than \(h^4\). This naturally gives rise to a decomposition of
  \([0, 1]\) as \(\bigcup_{1 \leq l \leq \lceil 1/h^4 \rceil}I_l\),
  where \(I_l = [v_{l - 1}, v_l]\).

  Note that
  \begin{align}
    &\quad \sup_{u \in [0, 1]}\left\|\sum_{\bc = 1}\bN_i(u)\right\|\nonumber\\
    &\leq \max_{1 \leq l \leq \lceil 1/h^4 \rceil} \left\|\sum_{\bc = 1}\bN_i(v_l)\right\| + \max_{1 \leq l \leq \lceil 1/h^4 \rceil} \sup_{u \in I_l} \left\|\sum_{\bc = 1}\bN_i(u) - \sum_{\bc = 1}\bN_i(v_l)\right\|\label{eq:s13}
  \end{align}

  The first term of~\eqref{eq:s13} shares the same upper bound
  \(\Delta_1\sqrt{\frac{r\log p}{hn_{\bc}}} +
  \Delta_1\frac{\sqrt{k}(\log p)^{3/2}}{hn_{\bc}}\) as
  in~\eqref{eq:s12} with probability exceeding
  \(1 - O(h^{-4}n_{\bc}p^{-11.5})\). For the second term
  of~\eqref{eq:s13}, note that
  \begin{align*}
    &\quad \sup_{u \in I_l} \left\|\sum_{\bc = 1}\bN_i(u) - \sum_{\bc = 1}\bN_i(v_l)\right\|\\
    &\leq \sup_{u \in I_l} \sum_{\bc = 1}|w_i(u) - w_i(v_l)|\| \bL(u_i) \btheta_i\|\\
    &\leq \sup_{u \in I_l} \sum_{\bc = 1}|w_i'(u_m)||u - v_l|\|\bL(u_i)\|\|\btheta_i\|\\
    &\leq \sup_{u \in I_l} \sum_{\bc = 1}\frac{1}{n_{\bc}h^2}\left|K'\left(\frac{u_i - u_m}{h}\right)\right|h^4\Delta_1 \sqrt{k}\|\btheta_i\|_{\infty}\\
    &\leq \frac{C_K h^2\Delta_1\sqrt{k}}{n_{\bc}} \sum_{\bc = 1}\|\btheta_i\|_{\infty},
  \end{align*}
  where the existence of \(u_m \in I_l\) is a consequence of the mean
  value theorem. Once again we use the fact
  \(\rP(\|\btheta_i\|_{\infty} \leq 5\sqrt{\log p}) \geq 1 -
  p^{-11.5}\). Combining it with the union bound, we get
  \(\rP(\sum_{\bc = 1}\|\btheta_i\|_{\infty} \leq 5n_{\bc}\sqrt{\log
    p}) \geq 1 - n_{\bc}p^{-11.5}\). Using the union bound again, we
  obtain that with probability exceeding
  \(1 - O(h^{-4}n_{\bc}p^{-11.5})\),
  \[\max_{1 \leq l \leq \lceil 1/h^4 \rceil} \sup_{u \in I_l} \left\|\sum_{\bc = 1}\bN_i(u) - \sum_{\bc = 1}\bN_i(v_l)\right\|
    \lesssim h^2\Delta_1\sqrt{k\log p}.\] This proves~\eqref{eq:s8} in
  view of~\eqref{eq:s13} and the union bound, and it also wraps up the
  proof of~\eqref{eq:s5} as previously discussed.
\end{myproof}

\begin{myproof}[Proof of~\eqref{eq:s6}]
  Note that
  \begin{align*}
    &\quad \sum_{\bc = 1} \frac{w_i}{w}\bx_{i}\bx_{i}^{\top} - \bPi^{(1)}\\
    &= \sum_{\bc = 1} \frac{w_i}{w}\bx_{i}\bx_{i}^{\top} - \bL(u)\bL(u)^{\top} - \sigma(u)^2\bI_p - \bmu^{(1)}(u)\bmu^{(1)}(u)^{\top}\\
    &= \sum_{\bc = 1} \frac{w_i}{w} \bL(u_i)\btheta_i\btheta_i^{\top}\bL(u_i)^{\top} - \bL(u)\bL(u)^{\top}\\
    &\quad + \sum_{\bc = 1} \frac{w_i}{w}\sigma(u_i)^2 \bEta_i \bEta_i^{\top} - \sigma(u)^2 \bI_p\\
    &\quad + \sum_{\bc = 1} \frac{w_i}{w}\bmu^{(1)}(u_i)\bmu^{(1)}(u_i)^{\top} - \bmu^{(1)}(u)\bmu^{(1)}(u)^{\top}\\
    &\quad + \sum_{\bc = 1} \frac{w_i}{w} \bL(u_i)\btheta_i\sigma(u_i)\bEta_i^{\top} + \sum_{\bc = 1}\frac{w_i}{w} \sigma(u_i)\bEta_i\btheta_i^{\top}\bL(u_i)^{\top}\\
    &\quad + \sum_{\bc = 1} \frac{w_i}{w} \bL(u_i)\btheta_i\bmu^{(1)}(u_i)^{\top} + \sum_{\bc = 1}\frac{w_i}{w} \bmu^{(1)}(u_i)\btheta_i^{\top}\bL(u_i)^{\top}\\
    &\quad + \sum_{\bc = 1} \frac{w_i}{w} \sigma(u_i)\bEta_i\bmu^{(1)}(u_i)^{\top} + \sum_{\bc = 1} \frac{w_i}{w} \bmu^{(1)}(u_i)\sigma(u_i)\bEta_i^{\top}.
  \end{align*}
  It suffices to prove the following six inequalities:
  \begin{multline}
    \label{eq:s14}
    \rP\Bigg(\sup_{u \in [0, 1]}\left\|\sum_{\bc = 1} \frac{w_i}{w} \bL(u_i)\btheta_i\btheta_i^{\top}\bL(u_i)^{\top} - \bL(u)\bL(u)^{\top}\right\|\\
    \gtrsim \Delta_1^2\sqrt{\frac{k\log p}{hn_{\bc}}} + \Delta_1^2\frac{k(\log p)^2}{hn_{\bc}} + h^2\Delta_1^2kp\Bigg) \lesssim h^{-4}n_{\bc}p^{-11.5},
  \end{multline}
  \begin{multline*}
    \rP\Bigg(\sup_{u \in [0, 1]}\left\|\sum_{\bc = 1}\frac{w_i}{w}\sigma(u_i)^2\bEta_i \bEta_i^{\top} - \sigma(u)^2 \bI_p\right\|\\
    \gtrsim \gamma^2\sqrt{\frac{p\log p}{hn_{\bc}}} + \gamma^2\frac{p(\log p)^2}{hn_{\bc}} + h^2\gamma^2 p \log p\Bigg) \lesssim h^{-4}n_{\bc}p^{-11.5},
  \end{multline*}
  \begin{multline*}
    \rP\Bigg(\sup_{u \in [0, 1]}\left\|\sum_{\bc = 1} \frac{w_i}{w}\bmu^{(1)}(u_i)\bmu^{(1)}(u_i)^{\top} - \bmu^{(1)}(u)\bmu^{(1)}(u)^{\top}\right\|\\
    \gtrsim M^2\sqrt{\frac{\log p}{hn_{\bc}}} + M^2\frac{\log p}{hn_{\bc}} + h^2M^2p\Bigg) \lesssim h^{-4}p^{-11.5}
  \end{multline*}
  \begin{multline*}
    \rP\Bigg(\sup_{u \in [0, 1]}\left\|\sum_{\bc = 1} \frac{w_i}{w} \bL(u_i)\btheta_i\sigma(u_i)\bEta_i^{\top}\right\|\\
    \gtrsim \Delta_1\gamma\sqrt{\frac{p\log p}{hn_{\bc}}} + \Delta_1\gamma\frac{\sqrt{kp}(\log p)^2}{hn_{\bc}} + h^2\Delta_1\gamma \sqrt{kp}\log p\Bigg) \lesssim h^{-4}n_{\bc}p^{-11.5}
  \end{multline*}
  \begin{multline*}
    \rP\Bigg(\sup_{u \in [0, 1]}\left\|\sum_{\bc = 1} \frac{w_i}{w} \bL(u_i)\btheta_i\bmu^{(1)}(u_i)^{\top}\right\|\\
    \gtrsim \Delta_1M\sqrt{\frac{k\log p}{hn_{\bc}}} + \Delta_1M\frac{\sqrt{k}(\log p)^{3/2}}{hn_{\bc}} + h^2\Delta_1M \sqrt{k\log p}\Bigg) \lesssim h^{-4}n_{\bc}p^{-11.5}
  \end{multline*}
  \begin{multline*}
    \rP\Bigg(\sup_{u \in [0, 1]}\left\|\sum_{\bc = 1} \frac{w_i}{w} \sigma(u_i)\bEta_i\bmu^{(1)}(u_i)^{\top}\right\|\\
    \gtrsim \gamma M\sqrt{\frac{p\log p}{hn_{\bc}}} + \gamma M\frac{\sqrt{p}(\log p)^{3/2}}{hn_{\bc}} + h^2\gamma M \sqrt{p\log p}\Bigg) \lesssim h^{-4}n_{\bc}p^{-11.5}
  \end{multline*}
  We only prove~\eqref{eq:s14} here as the rest are similar.

  We claim that (with the proofs deferred to improve readability)
  \begin{multline}
    \label{eq:s15}
    \rP\Bigg(\sup_{u \in [0, 1]}\left\|\sum_{\bc = 1} \bM_i - \sum_{\bc = 1} \E \bM_i\right\|\\
    \gtrsim \Delta_1^2\sqrt{\frac{k\log p}{hn_{\bc}}} + \Delta_1^2\frac{k(\log p)^2}{hn_{\bc}} + h^2\Delta_1^2k\log p\Bigg) \lesssim h^{-4}n_{\bc}p^{-11.5},
  \end{multline}
  and
  \begin{equation}
    \label{eq:s16}
    \sup_{u \in [0, 1]}\left\|\sum_{\bc = 1} \E \bM_i - \bL(u)\bL(u)^{\top}f_U(u)\right\| \lesssim h^2\Delta_1^2kp,
  \end{equation}
  where \(\bM_i(u)\) is defined as
  \(\bM_i(u) = w_i(u) \bL(u_i) \btheta_i \btheta_i^{\top}
  \bL(u_i)^{\top}\).

  Combining~\eqref{eq:s15} and~\eqref{eq:s16}, we get
  \begin{equation}
    \label{eq:s17}
    \rP\Bigg(\sup_{u \in [0, 1]}\left\|\sum_{\bc = 1} \bM_i - \bL(u)\bL(u)^{\top}f_U(u)\right\| \gtrsim \tilde{B} \Bigg) \lesssim h^{-4}n_{\bc}p^{-11.5},
  \end{equation}
  where
  \(\tilde{B} = \Delta_1^2\sqrt{\frac{k\log p}{hn_{\bc}}} +
  \Delta_1^2\frac{k(\log p)^2}{hn_{\bc}} + h^2\Delta_1^2kp\).
  
  We consider the event \(\cE = \cE_1 \cap \cE_2\), where
  \(\cE_1 \coloneq\{\sup_{u \in [0, 1]}|w - f_U(u)| \lesssim
  \sqrt{\frac{\log p}{hn_{\bc}}} + h^2\}\) and
  \(\cE_2 \coloneq \{\sup_{u \in [0, 1]}\|\sum_{\bc = 1} \bM_i -
  \bL(u)\bL(u)^{\top}f_U(u)\| \lesssim \tilde{B}\}\). Apply the union
  bound to Lemma~\ref{lem:s4} and~\eqref{eq:s17}, we got
  \(\rP(\cE^{c}) \lesssim h^{-4}n_{\bc}p^{-11.5}\). Since
  \(f_U \geq C_U > 0\) on \([0, 1]\), we have \(w \gtrsim C_U/2\)
  under \(\cE\). As a result, under \(\cE\),
  \begin{align*}
    &\quad \left\|\sum_{\bc = 1} \frac{w_i}{w} \bL(u_i)\btheta_i\btheta_i^{\top}\bL(u_i)^{\top} - \bL(u)\bL(u)^{\top}\right\|\\
    &=\left\|\frac{\sum_{\bc = 1}\bM_i}{w} - \frac{\bL(u)\bL(u)^{\top}f_U(u)}{f_U(u)}\right\|\\
    &\leq \frac{|f_U(u)|\left\|\sum_{\bc = 1} \bM_i - \bL(u)\bL(u)^{\top}f_U(u)\right\| + \left\|\bL(u)\bL(u)^{\top}f_U(u)\right\|\left|w - f_U(u)\right|}{|wf_U(u)|}\\
    &\lesssim \frac{\tilde C_U \tilde B + \Delta_1^2 \tilde C_U (\sqrt{(\log p)/(hn_{\bc})} + h^2)}{C_U^2/2}\\
    &\lesssim \tilde B.
  \end{align*}
  This proves~\eqref{eq:s14} and wraps up the proof of~\eqref{eq:s6}
  as previously discussed.
\end{myproof}

\begin{myproof}[Proof of~\eqref{eq:s15}]
  A two-step procedure is employed, where we first derive an upper
  bound of \(\|\sum_{\bc = 1}\bM_i(u) - \sum_{\bc = 1}\E\bM_i(u)\|\)
  for each fixed \(u\) using Lemma~\ref{lem:s2}, and then extrapolate
  the result to all \(u \in [0, 1]\) using a grid-based argument.

  Step 1, assume \(u \in [0, 1]\) is fixed. Since
  \(\bM_i = w_i \bL(u_i) \btheta_i \btheta_i^{\top} \bL(u_i)^{\top}\)
  where \(w_i = K_h(u_i - u)/n_{\bc}\), we have
  \(\|\bM_i\| \leq |w_i|\|\bL(u_i)\|^2\|\btheta_i\|^2 \leq
  \frac{C_K\Delta_1^2}{n_{\bc}h} \|\btheta_i\|^2 \leq
  \frac{C_K\Delta_1^2k}{n_{\bc}h} \|\btheta_i\|_{\infty}^2\) by
  Assumption~\ref{asp:3} and Assumption~\ref{asp:6}. Also note that
  \(\|\E[\bM_i|u_i]\| = \|w_i\bL(u_i)\bL(u_i)^{\top}\| \leq
  \frac{C_K\Delta_1^2k}{n_{\bc}h}\), thus
  \(\|\E\bM_i\| = \|\E\{\E[\bM_i|u_i]\}\| \leq \E\|\E[\bM_i|u_i]\|
  \leq \frac{C_K\Delta_1^2 k}{n_{\bc}h}\). By the normality of
  \(\btheta_i\), we can easily verify
  \(\rP(\|\btheta_i\|_{\infty} \leq 5\sqrt{\log p}) \geq 1 -
  p^{-11.5}\). As a result,
  \[\rP\left(\|\bM_i - \E\bM_i\| \geq \frac{C_K\Delta_1^2 k}{n_{\bc}h}(1 + 25 \log
      p)\right) \leq p^{-11.5}.\] To apply Lemma~\ref{lem:s2}, we
  define
  \(L \coloneq \frac{C_K\Delta_1^2 (k + 2)}{n_{\bc}h}(1 + 25\log p)\)
  and \(q_0 \coloneq p^{-11.5}\). Next we want to find \(q_1\).

  First, note that
  \(\E[\bM_i 1_{\|\btheta_i\|^2 \geq \tilde{L}}] - \E[\bM_i
  1_{\|\bM_i\| \geq L}]\) is positive semi-definite, where
  \(\tilde{L} = Ln_{\bc}h/(C_K\Delta_1^2) = (k + 2)(1 + 25\log p)\),
  so we have
  \begin{align*}
    &\quad \|\E[\bM_i 1_{\|\bM_i\| \geq L}]\| \leq \|\E[\bM_i 1_{\|\btheta_i\|^2 \geq \tilde{L}}]\| \leq \E\{\|\E[\bM_i 1_{\|\btheta_i\|^2 \geq \tilde{L}}|u_i]\|\}\\
    &= \E\{\|w_i\bL(u_i)\E[\btheta_i\btheta_i^{\top} 1_{\|\btheta_i\|^2 \geq \tilde{L}}]\bL(u_i)^{\top}\|\}\\
    &\leq \frac{C_K\Delta_1^2}{n_{\bc}h}\|\E[\btheta_i\btheta_i^{\top} 1_{\|\btheta_i\|^2 \geq \tilde{L}}]\|.
  \end{align*}
  By symmetric properties of the Gaussian distribution,
  \(\E[\btheta_i\btheta_i^{\top} 1_{\|\btheta_i\|^2 \geq \tilde{L}}]\)
  is a multiple of the identity matrix, thus its spectral norm is
  \(\E[\theta_{i1}^2 1_{\|\btheta_i\|^2 \geq \tilde{L}}] =
  \frac1{n_{\bc}}\E[\|\btheta_i\|^2 1_{\|\btheta_i\|^2 \geq
    \tilde{L}}]\). Since \(\|\btheta_i\|^2 \sim \chi^2(k)\), we have
  \begin{align*}
    \E[\|\btheta_i\|^2 1_{\|\btheta_i\|^2 \geq \tilde{L}}] &= \int_{\tilde{L}}^{\infty} x\frac{x^{k/2 - 1}e^{-x/2}}{2^{k/2}\Gamma(k/2)} dx\\
                                                           &= \frac{2\Gamma(k/2 + 1)}{\Gamma(k/2)} \int_{\tilde{L}}^{\infty} \frac{x^{(k + 2)/2 - 1}e^{-x/2}}{2^{(k + 2)/2}\Gamma((k + 2)/2)} dx\\
                                                           &= k \rP(Q \geq \tilde{L})
  \end{align*}
  where \(Q \sim \chi^2(k + 2)\). The Laurent-Massart bound for \(Q\)
  is \(\rP(Q \geq (k + 2) + 2\sqrt{(k + 2)x} + 2x) \leq \exp(-x)\).
  Let \(x = \frac{25}{4}(k + 2)\log p\), we can easily verify
  \(P(Q \geq \tilde{L}) \leq p^{-25k/4}\). As a result
  \[\|\E[\bM_i 1_{\|\bM_i\| \geq L}]\| \leq
    \frac{C_K\Delta_1^2k}{n_{\bc}^2h}p^{-\frac{25}{4}k}.\] We define
  \(q_1 \coloneq \frac{C_K\Delta_1^2k}{n_{\bc}^2h}p^{-25k/4}\).

  To find the matrix variance statistic \(V\), first note that
  \(\E[\bM_i^2] - \E[(\bM_i - \E\bM_i)^2]\) and
  \(\E\Delta_1^2w_i^2\|\btheta_i\|^2\bL(u_i)\btheta_i\btheta_i^{\top}\bL(u_i)^{\top}
  - \E[\bM_i^2]\) are positive semi-definite, so
  \begin{align*}
    V &= n_{\bc}\|\E[(\bM_i - \E\bM_i)^2]\|\\
      &\leq n_{\bc}\|\E[\bM_i^2]\| \leq n_{\bc}\Delta_1^2\left\|\E[w_i^2\|\btheta_i\|^2\bL(u_i)\btheta_i\btheta_i^{\top}\bL(u_i)^{\top}]\right\|\\
      &\leq n_{\bc}\Delta_1^2\E\left\{\left\|\E[w_i^2\|\btheta_i\|^2\bL(u_i)\btheta_i\btheta_i^{\top}\bL(u_i)^{\top}|u_i]\right\|\right\}\\
      &= n_{\bc}\Delta_1^2 \E\left\{w_i^2\left\|\bL(u_i)\E[\|\btheta_i\|^2\btheta_i\btheta_i^{\top}]\bL(u_i)^{\top}\right\|\right\}\\
      &\leq n_{\bc}\Delta_1^4\left\|\E[\|\btheta_i\|^2\btheta_i\btheta_i^{\top}]\right\| \E[w_i^2] \leq \frac{\tilde C_U C_K \Delta_1^4}{n_{\bc}h}\E[\|\btheta_i\|^2\btheta_i\btheta_i^{\top}],
  \end{align*}
  where the last inequality is due to~\eqref{eq:s11}. By symmetric
  properties of the Gaussian distribution,
  \(\E[\|\btheta_i\|^2\btheta_i\btheta_i^{\top}]\) is a multiple of
  the identity matrix, thus its spectral norm is
  \(\E[\|\btheta_i\|^2 \theta_{i1}^2] = \E[\theta_{i1}^4] + \sum_{j =
    2}^k\E[\theta_{ij}^2]\E[\theta_{i1}^2] = k + 2\). As a result,
  \[V \leq \frac{\tilde C_U C_K\Delta_1^4}{n_{\bc}h}(k + 2).\]

  Now we can apply Lemma~\ref{lem:s2} and conclude with probability
  exceeding \(1-O(n_{\bc}p^{-11.5})\)
  \begin{equation}
    \label{eq:s18}
    \left\|\sum_{\bc = 1}\bM_i(u) - \sum_{\bc = 1}\E\bM_i(u)\right\| \lesssim
    \Delta_1^2\sqrt{\frac{k\log p}{hn_{\bc}}} +
    \Delta_1^2\frac{k(\log p)^2}{hn_{\bc}}.
  \end{equation}

  Step 2, we construct a equally spaced grid in \([0, 1]\) with
  \(\lceil 1/h^4\rceil + 1\) grid points
  \(\{v_l: 0 \leq l \leq \lceil 1/h^4 \rceil\}\) (including end
  points). The distance between neighboring grid points is thus less
  than \(h^4\). This naturally gives rise to a decomposition of
  \([0, 1]\) as \(\bigcup_{1 \leq l \leq \lceil 1/h^4 \rceil}I_l\),
  where \(I_l = [v_{l - 1}, v_l]\).

  Note that
  \begin{align}
    &\quad \sup_{u \in [0, 1]}\left\|\sum_{\bc = 1}\bM_i(u) - \sum_{\bc = 1}\E\bM_i(u)\right\|\nonumber\\
    &\leq \max_{1 \leq l \leq \lceil 1/h^4 \rceil} \left\|\sum_{\bc = 1}\bM_i(v_l) - \sum_{\bc = 1}\E\bM_i(v_l)\right\|\nonumber\\
    &\quad + \max_{1 \leq l \leq \lceil 1/h^4 \rceil} \sup_{u \in I_l} \left\|\left(\sum_{\bc = 1}\bM_i(u) - \sum_{\bc = 1}\E\bM_i(u)\right) - \left(\sum_{\bc = 1}\bM_i(v_l) - \sum_{\bc = 1}\E\bM_i(v_l)\right)\right\|\nonumber\\
    &\leq \max_{1 \leq l \leq \lceil 1/h^4 \rceil} \left\|\sum_{\bc = 1}\bM_i(v_l) - \sum_{\bc = 1}\E\bM_i(v_l)\right\|\label{eq:s19}\\
    &\quad + \max_{1 \leq l \leq \lceil 1/h^4 \rceil} \sup_{u \in I_l} \left\|\sum_{\bc = 1}\bM_i(u) - \sum_{\bc = 1}\bM_i(v_l)\right\|\label{eq:s20}\\
    &\quad + \max_{1 \leq l \leq \lceil 1/h^4 \rceil} \sup_{u \in I_l} \left\|\sum_{\bc = 1}\E\bM_i(u) - \sum_{\bc = 1}\E\bM_i(v_l)\right\|\label{eq:s21}
  \end{align}

  Using the union bound, \eqref{eq:s19} shares the same upper bound
  \(\Delta_1^2\sqrt{\frac{k\log p}{hn_{\bc}}} + \Delta_1^2\frac{k(\log
    p)^2}{hn_{\bc}}\) as in~\eqref{eq:s18} with probability exceeding
  \(1 - O(h^{-4}n_{\bc}p^{-11.5})\). For the term~\eqref{eq:s20}, note
  that
  \begin{align*}
    &\quad \sup_{u \in I_l} \left\|\sum_{\bc = 1}\bM_i(u) - \sum_{\bc = 1}\bM_i(v_l)\right\|\\
    &\leq \sup_{u \in I_l} \sum_{\bc = 1}|w_i(u) - w_i(v_l)|\| \bL(u_i) \btheta_i \btheta_i^{\top} \bL(u_i)^{\top}\|\\
    &\leq \sup_{u \in I_l} \sum_{\bc = 1}|w_i'(u_m)||u - v_l|\|\bL(u_i)\|^2\|\btheta_i\|^2\\
    &\leq \sup_{u \in I_l} \sum_{\bc = 1}\frac{1}{n_{\bc}h^2}\left|K'\left(\frac{u_i - u_m}{h}\right)\right|h^4\Delta_1^2 k\|\btheta_i\|_{\infty}^2\\
    &\leq \frac{C_K h^2\Delta_1^2k}{n_{\bc}} \sum_{\bc = 1}\|\btheta_i\|_{\infty}^2,
  \end{align*}
  where the existence of \(u_m \in I_l\) is a consequence of the mean
  value theorem. Once again we use the fact
  \(\rP(\|\btheta_i\|_{\infty} \leq 5\sqrt{\log p}) \geq 1 -
  p^{-11.5}\). Combining it with the union bound, we get
  \(\rP(\sum_{\bc = 1}\|\btheta_i\|_{\infty}^2 \leq 25n_{\bc}\log p)
  \geq 1 - n_{\bc}p^{-11.5}\). Using the union bound again, we obtain
  that with probability exceeding \(1 - O(h^{-4}n_{\bc}p^{-11.5})\),
  \[\max_{1 \leq l \leq \lceil 1/h^4 \rceil} \sup_{u \in I_l} \left\|\sum_{\bc = 1}\bM_i(u) - \sum_{\bc = 1}\bM_i(v_l)\right\|
    \lesssim h^2\Delta_1^2k\log p.\] For the term~\eqref{eq:s21}, we
  have the following similar derivation:
  \begin{align*}
    &\quad \sup_{u \in I_l} \left\|\sum_{\bc = 1}\E\bM_i(u) - \sum_{\bc = 1}\E\bM_i(v_l)\right\|\\
    &\leq \sup_{u \in I_l} \sum_{\bc = 1} \|\E\{\E[\bM_i(u) - \bM_i(v_l)|u_i]\}\|\\
    &= \sup_{u \in I_l} \sum_{\bc = 1}\|\E\{(w_i(u) - w_i(v_l))\bL(u_i)\E[\btheta_i \btheta_i^{\top}]\bL(u_i)^{\top}\}\|\\
    &\leq \sup_{u \in I_l} \sum_{\bc = 1}\E[|w_i(u) - w_i(v_l)|\|\bL(u_i)\|^2]\\
    &\leq \sup_{u \in I_l} \sum_{\bc = 1}\E\left[\frac{1}{n_{\bc}h^2}\left|K'\left(\frac{u_i - u_m}{h}\right)\right|h^4\Delta_1^2\right]\\
    &\lesssim h^2\Delta_1^2.
  \end{align*}
  Combining the bounds for terms~\eqref{eq:s19}--\eqref{eq:s21} using
  the union bound, we finish the proof of~\eqref{eq:s15}.
\end{myproof}

\begin{myproof}[Proof of~\eqref{eq:s16}]
  \begin{align*}
    \sum_{\bc = 1} \E \bM_i &= \sum_{\bc = 1} \E\{\E[w_i\bL(u_i)\btheta_i\btheta_i^{\top}\bL(u_i)^{\top}|u_i]\}\\
                            &=\sum_{\bc = 1} \E \{w_i\bL(u_i)\E[\btheta_i\btheta_i^{\top}]\bL(u_i)^{\top}\}\\
                            &=\sum_{\bc = 1} \E [w_i\bL(u_i)\bL(u_i)^{\top}]\\
                            &=\sum_{\bc = 1} \int_{\R}\frac{1}{n_{\bc}h}K\left(\frac{v - u}{h}\right)\bL(v)\bL(v)^{\top}f_U(v)dv\\
                            &=\int_{\R}K(\nu)\bL(u + h\nu)\bL(u + h\nu)^{\top}f_U(u + h\nu)d\nu\\
                            &\overset{\mathrm{(i)}}{=}\int_{\R}K(\nu)\left(\bL(u) + h\nu\bL'(u) + \frac{h^2\nu^2}{2}\bL''_m\right)\\
                            &\quad \cdot\left(\bL(u) + h\nu\bL'(u) + \frac{h^2\nu^2}{2}\bL''_m\right)^{\top}\left(f_U(u) + h\nu f'_U(u) + \frac{h^2\nu^2}{2}f''_U(u_m)\right)d\nu\\
                            &\overset{\mathrm{(ii)}}{=}\bL(u)\bL(u)^{\top}f_U(u)\\
                            &\quad + \int_{\R}K(\nu)\frac{h^2\nu^2}{2}(\bL''_m \bL(u)^\top f_U(u) + \bL(u)\bL''^{\top}_m f_U(u) + \bL(u)\bL(u)^{\top}f''_U(u))d\nu\\
                            &\quad + \int_{\R}K(\nu)h^2\nu^2(\bL(u)\bL'(u)^{\top}f'_U(u) + \bL'(u)\bL(u)^{\top}f'_U(u) + \bL'(u)\bL'(u)^{\top}f_U(u))d\nu\\
                            &\quad + \int_{\R}K(\nu)\frac{h^4\nu^4}{4}(\bL(u)\bL''^{\top}_m f''_U(u_m) + \bL''_m \bL(u)^{\top}f''_U(u_m) + \bL''_m \bL''^{\top}_m f_U(u_m))d\nu\\
                            &\quad + \int_{\R}K(\nu)\frac{h^4\nu^4}{2}(\bL''_m\bL'(u)^{\top}f'_U(u) + \bL'(u)\bL''^{\top}_m f'_U(u) + \bL'(u)\bL'(u)^{\top}f''_U(u))d\nu\\
                            &\quad + \int_{\R}K(\nu)\frac{h^6\nu^6}{8}\bL''_m\bL''^{\top}_m f''_U(u_m)d\nu.
  \end{align*}
  where (i) applies Taylor's theorem (with the remainders taking the
  mean-value form), with \((\bL''_m)_{jl} = \bL''_{jl}(u_{jl})\),
  \(u_{jl} \in (u, u + h\nu)\) and \(u_m \in (u, u + h\nu)\), and (ii)
  uses Assumption~\ref{asp:6}.

  As a result, using Assumption~\ref{asp:3},~\ref{asp:1}
  and~\ref{asp:6}, we have
  \begin{align*}
    &\quad\left\|\sum_{\bc = 1} \E \bM_i - \bL(u)\bL(u)^{\top}f_U(u)\right\|\\
    &\lesssim h^2 \int_{\R}K(\nu)\nu^2\|\bL(u)\|^2|f_U''(u)|d\nu\\
    &\quad + h^2 \int_{\R}K(\nu)\nu^2\sqrt{kp}(\|\bL''_m\|_{\infty} \|\bL(u)\||f_U(u)| + \|\bL(u)\|\|\bL'(u)\|_{\infty}|f'_U(u)|)d\nu\\
    &\quad + h^2 \int_{\R}K(\nu)\nu^2kp\|\bL'(u)\|_{\infty}^2|f_U(u)|d\nu\\
    &\quad + h^4 \int_{\R}K(\nu)\nu^4\sqrt{kp}\|\bL''_m\|_{\infty}\|\bL(u)\||f''_U(u)|d\nu\\
    &\quad + h^4 \int_{\R}K(\nu)\nu^4kp(\|\bL''_m\|_{\infty}^2|f_U(u)| + \|\bL''_m\|_{\infty}\|\bL'(u)\|_{\infty}|f'_U(u)| + \|\bL'(u)\|^2_{\infty}|f''_U(u)|)d\nu\\
    &\quad + h^6 \int_{\R}K(\nu)\nu^6kp\|\bL''_m\|_{\infty}^2|f''_U(u)|d\nu\\
    &\lesssim h^2\Delta_1^2 + h^2\Delta_1^2\sqrt{kp} + h^2\Delta_1^2kp + h^4\Delta_1^2\sqrt{kp} + h^4\Delta_1^2kp + h^6\Delta_1^2kp\\
    &\lesssim h^2\Delta_1^2kp.
  \end{align*}
  The last inequality holds because \(h \to 0\). Since inequality
  above is true for all \(u \in [0, 1]\), we have completed the proof
  of~\eqref{eq:s16}.
\end{myproof}

\begin{lemma}
  \label{lem:s5}
  Under Assumption~\ref{asp:5},
  \(\lambda_{k + 1}(\bSigma^{tot}_{\rho}) - \lambda_{k +
    2}(\bSigma^{tot}_{\rho}) \geq C_{\phi} (\Delta_k^2 \wedge
  m^2)/2\).
\end{lemma}
\begin{myproof}
  First, we decompose \(\R^p\) as the sum of \(W_1\) and \(W_2\),
  where \(W_1\) is spanned by the columns of \(\bL\) and \(\bdelta\),
  and \(W_2\) is the orthogonal complement of \(W_1\). By
  Assumption~\ref{asp:5}, we have \(\dim(W_1) = k + 1\).

  For any \(\bbv \in W_2\),
  \(\bSigma^{tot}_{\rho}\bbv = (\bL\bL^{\top} + \sigma^2\bI_p + \rho
  \bdelta\bdelta^{\top})\bbv = \sigma^2\bbv\). Thus \(W_1\) and
  \(W_2\) are invariant subspaces of the symmetric matrix
  \(\bSigma^{tot}_{\rho}\). Next, we will show
  \begin{equation}
    \label{eq:s22}
    \bbv^{\top} (\bL\bL^{\top} + \rho \bdelta\bdelta^{\top}) \bbv
    \geq \frac{C_{\phi}}{2} (\Delta_k^2 \wedge m^2)
  \end{equation}
  for any unit vector \(\bbv \in W_1\), which implies
  \(\lambda_{k + 1}(\bSigma^{tot}_{\rho}) \geq \sigma^2 + C_{\phi}
  (\Delta_k^2 \wedge m^2)/2\) and
  \(\lambda_{k + 2}(\bSigma^{tot}_{\rho}) = \sigma^2\). This
  immediately completes the proof of the lemma.

  Recall that \(0 < \phi \leq \pi/2\) is the angle between \(\bdelta\)
  and \(C(\bL)\), where \(C(\bL)\) is the column space of \(\bL\). Let
  \(0 \leq \psi \leq \pi/2\) denote the angle between the unit vector
  \(\bbv \in W_1\) and the one-dimensional subspace
  \(\R\bdelta_{\perp}\), where
  \(\bdelta_{\perp} = \bdelta - \bdelta_{\parallel}\) and
  \(\bdelta_{\parallel}\) is the projection of \(\bdelta\) onto
  \(C(\bL)\). Also let \(\tilde \bdelta = \bdelta/\|\bdelta\|\),
  \(\tilde\rho = \rho \|\bdelta\|^2\) and
  \(\tilde \bSigma = \bL\bL^{\top} + \rho \bdelta\bdelta^{\top} =
  \bL\bL^{\top} + \tilde \rho \tilde \bdelta \tilde \bdelta^{\top}\).
  Note that \(\tilde \rho \geq m^2\) by the definition of \(m\)
  following Assumption~\ref{asp:5}. With fixed \(\phi\) and varying
  \(\psi\), we prove \eqref{eq:s22} by two cases:
  \begin{enumerate}
  \item \(\psi \geq \phi\). Let \(\bbv_{\parallel}\) denote the
    projection of \(\bbv\) onto \(C(\bL)\). Then
    \(\bbv^{\top} \tilde \bSigma \bbv \geq \bbv_{\parallel}^{\top}
    \bL\bL^{\top} \bbv_{\parallel} \geq
    \|\bbv_{\parallel}\|^2(\gamma_k - \delta^2) \geq \sin^2(\psi)
    \Delta_k^2 \geq C_{\phi} \Delta_k^2\).
  \item \(\psi < \phi\). Let \(0 \leq \omega \leq \pi/2\) denote the
    angle between \(\R\bbv\) and \(\R\bdelta\). Then
    \(\bbv^{\top} \tilde \bSigma \bbv \geq \sin^2(\psi) \Delta_k^2 +
    \cos^2(\omega)\tilde\rho\), the RHS of which attains its lower
    bound when \(\omega\) is maximized. However, the condition
    \(\psi < \phi\) ensures \(\omega < \pi/2\), so \(\omega\) is
    maximized when (1)\ \(\bbv\) lies in the space spanned by
    \(\bdelta\) and \(\bdelta_{\perp}\) and (2)\ \(\bbv\) is on the
    opposite side of \(\bdelta\) with respect to \(\bdelta_{\perp}\).
    That is, \(\omega = \pi/2 - \phi + \psi\). As a result,
    \begin{align*}
      \bbv^{\top} \tilde \bSigma \bbv &\geq \sin^2(\psi) \Delta_k^2 + \cos^2(\pi/2 - \phi + \psi) \tilde\rho\\
                                      &= \frac{1 - \cos 2\psi}{2} \Delta_k^2 + \frac{1 - \cos(2\phi - 2\psi)}{2}\tilde\rho\\
                                      &= \frac{1}{2}(\Delta_k^2 + \tilde\rho) - \frac{1}{2}\left((\Delta_k^2 + \tilde \rho \cos 2\phi)\cos 2\psi + \tilde \rho \sin 2\phi \sin 2\psi\right)\\
                                      & \geq \frac{1}{2}(\Delta_k^2 + \tilde\rho) - \frac{1}{2} \sqrt{(\Delta_k^2 + \tilde \rho \cos 2\phi)^2 + \tilde \rho^2 \sin^2 2\phi}\\
                                      &= \frac{1}{2}(\Delta_k^2 + \tilde\rho) - \frac{1}{2}\sqrt{\Delta_k^4 + 2\Delta_k^2\tilde \rho (1 - 2\sin^2 \phi) + \tilde \rho^2}\\
                                      & \geq \frac{1}{2}(\Delta_k^2 + \tilde\rho) - \frac{1}{2}\sqrt{(\Delta_k^2 + \tilde \rho)^2 - 4C_{\phi}\Delta_k^2\tilde \rho}\\
                                      &=\frac{1}{2}(\Delta_k^2 + \tilde \rho) \left(1 - \sqrt{1 - \frac{4C_{\phi}\Delta_k^2 \tilde \rho}{(\Delta_k^2 + \tilde \rho)^2}}\right)\\
                                      & \geq \frac{1}{2}(\Delta_k^2 + \tilde \rho)\frac{2C_{\phi}\Delta_k^2 \tilde \rho}{(\Delta_k^2 + \tilde \rho)^2} = \frac{C_{\phi}\Delta_k^2 \tilde \rho}{\Delta_k^2 + \tilde \rho}\\
                                      & \geq \frac{C_{\phi}}{2}(\Delta_k^2 \wedge \tilde\rho) \geq \frac{C_{\phi}}{2}(\Delta_k^2 \wedge m^2).
    \end{align*}
  \end{enumerate}
\end{myproof}

\begin{myproof}[Proof of Theorem~\ref{thm:2}]
  Consider the event
  \(\cE_n = \{\sup_{u \in [0, 1]}\|\hat{\bSigma}_{\rho}^{tot}(u) -
  \bSigma_{\rho}^{tot}(u)\| \lesssim B_n\}\), where
  \(B_n = B_{\RN{1}}M + B_{\RN{1}}^2 + B_{\RN{2}}\). Under \(\cE_n\),
  we have
  \[\|\hat{\bSigma}_{\rho}^{tot} - \bSigma_{\rho}^{tot}\|/(\lambda_{k
      + 1}(\bSigma^{tot}_{\rho}) - \lambda_{k +
      2}(\bSigma^{tot}_{\rho})) \lesssim B_n(\Delta_k^2 \wedge
    m^2)^{-1} \to 0\] by Lemma~\ref{lem:s5}. This implies for all
  sufficiently large \(n\), the conditions of Corollary 2.8
  \citep{chen2021spectral} of the Davis-Kahan sin\(\bTheta\) Theorem
  are satisfied. As a result,
  \[d(\bR_1, \hat\bR_1) \lesssim \frac{\|\hat{\bSigma}_{\rho}^{tot} -
      \bSigma_{\rho}^{tot}\|}{\lambda_{k + 1}(\bSigma^{tot}_{\rho}) -
      \lambda_{k + 2}(\bSigma^{tot}_{\rho})} \lesssim B_n(\Delta_k^2
    \wedge m^2)^{-1}.\] Since this is true for arbitrary
  \(u \in [0, 1]\), we have
  \(\sup_{u \in [0, 1]}d(\bR_1, \hat\bR_1) \lesssim B_n(\Delta_k^2
  \wedge m^2)^{-1}\), and it happens under \(\cE_n\) with probability
  exceeding \(1 - O(h^{-4}np^{-11.5})\) by Lemma~\ref{lem:s3}.
\end{myproof}

\begin{myproof}[Proof of Corollary~\ref{cor:1}]
  Let \(\tilde{B}_{\RN{1}} = B_{\RN{1}}(\Delta_k^2 \wedge m^2)^{-1}\),
  \(\mathring{B}_{\RN{1}} = B_{\RN{1}}(\Delta_k^2 \wedge m^2)^{-1/2}\)
  and \(\tilde{B}_{\RN{2}} = B_{\RN{2}}(\Delta_k^2 \wedge m^2)^{-1}\).
  With \(h \asymp (\log p / (p^2n))^{1/5}\),
  \(\Delta_1^2 \vee M^2 \lesssim \Delta_k^2 \wedge m^2\),
  \(\gamma^2 \lesssim (\Delta_k^2 \wedge m^2)p^{-1}\), and
  \(\rho, r, k = O(1)\), we can easily verify that
  \(\tilde B_{\RN{1}}M \vee \mathring B_{\RN{1}} \lesssim B_0 +
  B_0^2(\log p)^{1/2}\) and
  \(\tilde B_{\RN{2}} \lesssim B_0 + B_0^2\log p\), where
  \(B_0 \coloneq p^{1/5}(\log p)^{2/5}n^{-2/5}\).

  Since \(p^{1/5}(\log p)^2 \lesssim n^{2/5}\), we have
  \(B_0\log p \to 0\), which also yields
  \(\tilde B_{\RN{1}}M \vee \mathring B_{\RN{1}}\vee \tilde B_{\RN{2}}
  \lesssim B_0 \to 0\). Thus
  \(B_n(\Delta_k^2 \wedge m^2)^{-1} = \tilde B_{\RN{1}}M + \mathring
  B_{\RN{1}}^2 + \tilde B_{\RN{2}} \lesssim B_0\). On the other hand,
  \(O(h^{-4}np^{-11.5}) \lesssim (\log p / (p^2n))^{-4/5}np^{-11.5}
  \lesssim n^2p^{-9}\). Applying these results to the conclusion of
  Theorem~\ref{thm:2} proves the corollary.
\end{myproof}

\begin{myproof}[Proof of Theorem~\ref{thm:3}]
  Define $a(u)=\lambda_{p}^{(1)}(u)=\lambda_{p}^{(2)}(u)$ and
  $a^{(c)}_{i}(u)=\lambda_{i}^{(c)}(u)-\lambda_{p}^{(c)}(u),c=1,2$. By
  the spiked assumption, we have $a^{(c)}_{i}(u)\geq 0$ for all $i$,
  and $a^{(c)}_{i}(u)=0$ for $i>k_c$. Let $\{\bnu_i^{(1)}(u)\}$ be
  eigenvectors of $\bSigma^{(1)}(u)$.
  \begin{align*}
    \bSigma^{(1)}(u)
    &=\sum_{i=1}^{p}\lambda_{i}^{(1)}(u)\bnu_{i}^{(1)}(u)(\bnu_{i}^{(1)}(u))^{\top} \\
    & = \sum_{i=1}^{p}(\lambda_{i}^{(1)}(u)-\lambda_{p}^{(1)}(u))\bnu_{i}^{(1)}(u)(\bnu_{i}^{(1)}(u))^{\top}+\lambda_{p}^{(1)}(u)\sum_{i=1}^{p}\bnu_{i}^{(1)}(u)(\bnu_{i}^{(1)}(u))^{\top}\\
    & = \sum_{i=1}^{k_{1}}a_{i}^{(1)}(u) \bnu_{i}^{(1)}(u)(\bnu_{i}^{(1)}(u))^{\top}+\lambda_{p}^{(1)}(u)\bI\\
    & =a(u)\bI+\sum_{i=1}^{k_{1}}a_{i}^{(1)}(u)\bnu_{i}^{(1)}(u)(\bnu_{i}^{(1)}(u))^{\top}
  \end{align*}
  Similarly,
  \begin{align*}
    \bSigma^{(2)}(u)& =a(u)\bI+\sum_{i=1}^{k_{2}}a^{(2)}_{i}(u)\bnu_{i}^{(2)}(u)(\bnu_{i}^{(2)}(u))^{\top}
  \end{align*}

  For easy presentation, we assume equal prior probabilities for both
  classes. Let
  $\bSigma(u)=\frac{\bSigma^{(1)}(u)+\bSigma^{(2)}(u)}{2}$. Then

  \begin{align*}
    \bSigma_{\rho}^{tot}(u)=&\bSigma(u)+\rho\bdelta(u)\bdelta^{\top}(u)\\
    =&a(u)\bI+\rho\bdelta(u)\bdelta^{\top}(u)\\&+\frac{1}{2}\sum_{i=1}^{k_{1}}(a^{(1)}_{i}(u)\bnu_{i}^{(1)}(u)(\bnu_{i}^{(1)}(u))^{\top})+\frac{1}{2}\sum_{i=1}^{k_{2}}(a^{(2)}_{i}(u)\bnu_{i}^{(2)}(u)(\bnu_{i}^{(2)}(u))^{\top}).
  \end{align*}

  Let $V\subset\R^p$ be a linear space spanned by
  $\{\bdelta^{\top}(u),\bnu_{1}^{(1)}(u),\dots,\bnu_{k_1}^{(1)}(u),\bnu_{1}^{(2)}(u),\dots,\bnu_{k_2}^{(2)}(u)\}$.
  In general, $\dim V=k_1+k_2+1$. Let $V^{\perp}$ be the orthogonal
  complement of $V$. It is straightforward to verify
  $\bSigma_{\rho}^{tot}(u)\bbv=a(u)\bbv$ for all $\bbv\in V^{\perp}$.
  This implies that $V$ is the space spanned by the eigenvectors of
  $\bSigma_{\rho}^{tot}(u)$ corresponding to its top $k_1+k_2+1$
  eigenvalues. Now we write $\bR(u)=(\bR_1(u),\bR_2(u))$ where
  $\bR_1(u)$ is a $p\times (k_1+k_2+1)$ matrix. The column space of
  $\bR_2(u)$ is $V^{\perp}$. Let
  \[\tilde\bx=\bR^{\top}(u)\bx=
    \begin{pmatrix}
      \bR_1^{\top}(u)\bx \\
      \bR_2^{\top}(u)\bx
    \end{pmatrix}
    =
    \begin{pmatrix}
      \tilde\bx_1 \\
      \tilde\bx_2
    \end{pmatrix}.\]
  Next, we will show the optimal QDA rule is independent of
  $\tilde\bx_2$. First of all, we have

  \begin{align*}
    &\bSigma^{(1)}(u)^{-1}=\left(a(u)\bI+\sum_{i=1}^{k_{1}}a^{(1)}_{i}(u)\bnu_{i}^{(1)}(u)(\bnu_{i}^{(1)}(u))^{\top}\right)^{-1}\\
    &= \frac{1}{a(u)}\bI-\sum_{i=1}^{k_{1}}\frac{a^{(1)}_{i}(u)}{a(u)(a(u)+a^{(1)}_{i}(u))}\bnu_{i}^{(1)}(u)(\bnu_{i}^{(1)}(u))^{\top}.
  \end{align*}

  This can be verified directly as follows.
  \begin{align*}
    &\left(a(u)\bI+\sum_{i=1}^{k_{1}}a^{(1)}_{i}(u)\bnu_{i}^{(1)}(u)(\bnu_{i}^{(1)}(u))^{\top}\right)\left(\frac{1}{a(u)}\bI-\sum_{i=1}^{k_{1}}\frac{a^{(1)}_{i}(u)}{a(u)(a(u)+a^{(1)}_{i}(u))}\bnu_{i}^{(1)}(u)(\bnu_{i}^{(1)}(u))^{\top}\right)\\
    &=\bI+\sum_{i=1}^{k_{1}}\frac{1}{a(u)}a^{(1)}_{i}(u)\bnu_{i}^{(1)}(u)(\bnu_{i}^{(1)}(u))^{\top}-
      \sum_{i=1}^{k_{1}}\frac{a^{(1)}_{i}(u)}{(a(u)+a^{(1)}_{i}(u))}\bnu_{i}^{(1)}(u)(\bnu_{i}^{(1)}(u))^{\top}\\
    &-\sum_{i=1}^{k_{1}}\frac{(a^{(1)}_{i}(u))^{2}}{a(u)(a(u)+a^{(1)}_{i}(u))}\bnu_{i}^{(1)}(u)(\bnu_{i}^{(1)}(u))^{\top}\\
    &=\bI.
  \end{align*}

  Similarly,
  \begin{align*}
    \bSigma^{(2)}(u)^{-1}=\frac{1}{a(u)}\bI-\sum_{i=1}^{k_{2}}\frac{a^{(2)}_{i}(u)}{a(u)(a(u)+a^{(2)}_{i}(u))}\bnu_{i}^{(2)}(u)(\bnu_{i}^{(2)}(u))^{\top}.
  \end{align*}

  Given a realization of $\bX=\bx,U=u$, the optimal QDA rule labels
  the observation to class 1 if
  \begin{align*}
    &\bx^{\top}(\bSigma^{(1)}(u)^{-1}-\bSigma^{(2)}(u)^{-1})\bx-2\bx^{\top}\left(\bSigma^{(1)}(u)^{-1}\bmu^{(1)}(u)-\bSigma^{(2)}(u)^{-1}\bmu^{(2)}(u)\right)\\
    &+(\bmu^{(1)}(u))^{\top}\bSigma^{(1)}(u)^{-1}\bmu^{(1)}(u)-(\bmu^{(2)}(u))^{\top}\bSigma^{(2)}(u)^{-1}\bmu^{(2)}(u)+\log{\frac{|\bSigma^{(1)}(u)|}{|\bSigma^{(2)}(u)|}}\geq 0.
  \end{align*}

  Therefore, we can write the discriminant function in a quadratic
  form
  \[\bx^{\top}\bA(u)\bx+2\bx^{\top}\bB(u)+C(u),\]
  where
  \begin{align*}
    \bA(u) &= \bSigma^{(1)}(u)^{-1}-\bSigma^{(2)}(u)^{-1}\\
                  &=\sum_{i=1}^{k_{2}}\frac{a^{(2)}_{i}(u)}{a(u)(a(u)+a^{(2)}_{i}(u))}\bnu_{i}^{(2)}(u)(\bnu_{i}^{(2)}(u))^{\top}\\
                  &\quad-\sum_{i=1}^{k_{1}}\frac{a^{(1)}_{i}(u)}{a(u)(a(u)+a^{(1)}_{i}(u))}\bnu_{i}^{(1)}(u)(\bnu_{i}^{(1)}(u))^{\top},
  \end{align*}

  \begin{align*}
    \bB(u)&=\bSigma^{(1)}(u)^{-1}\bmu^{(1)}(u)-\bSigma^{(2)}(u)^{-1}\bmu^{(2)}(u)\\
          &=\frac{1}{a(u)}\bmu^{(1)}(u)-\sum_{i=1}^{k_{1}}\frac{a^{(1)}_{i}(u)(\bnu_{i}^{(1)}(u))^{\top}\bmu^{(1)}(u)}{a(u)(a(u)+a^{(1)}_{i}(u))}\bnu_{i}^{(1)}(u)\\
          &-\frac{1}{a(u)}\bmu^{(2)}(u)+\sum_{i=1}^{k_{2}}\frac{a^{(2)}_{i}(u)(\bnu_{i}^{(2)}(u))^{\top}\bmu^{(2)}(u)}{a(u)(a(u)+a^{(2)}_{i}(u))}\bnu_{i}^{(2)}(u)\\
          &=\frac{1}{a(u)}\left(\bmu^{(1)}(u)-\bmu^{(2)}(u)\right)+\sum_{i=1}^{k_{2}}\frac{a^{(2)}_{i}(u)(\bnu_{i}^{(2)}(u))^{\top}\bmu^{(2)}(u)}{a(u)(a(u)+a^{(2)}_{i}(u))}\bnu_{i}^{(2)}(u)\\
          &-\sum_{i=1}^{k_{1}}\frac{a^{(1)}_{i}(u)(\bnu_{i}^{(1)}(u))^{\top}\bmu^{(1)}(u)}{a(u)(a(u)+a^{(1)}_{i}(u))}\bnu_{i}^{(1)}(u),
  \end{align*}

  \begin{align*}
    C(u)=(\bmu^{(1)}(u))^{\top}\bSigma^{(1)}(u)^{-1}\bmu^{(1)}(u)
    -(\bmu^{(2)}(u))^{\top}\bSigma^{(2)}(u)^{-1}\bmu^{(2)}(u)+ \log{\frac{|\bSigma_{1}(u)|}{|\bSigma_{2}(u)|}} .
  \end{align*}

  Now we verify the quadratic form depends only on the first
  $k_1+k_2+1$ coordinates of $\tilde{\bx}$.
  \begin{align*}
    &\bx^{\top}\bA(u)\bx\\
    =& (\bR(u)\bR^{\top}(u)\bx)^{\top}\bA(u)\bR(u)\bR^{\top}(u)\bx\\
    =&(\bR(u)\tilde\bx)^{\top}\bA(u)\bR(u)\tilde\bx\\
    =&\left((\bR_1(u),\bR_2(u))
       \begin{pmatrix}
         \tilde\bx_1 \\
         \tilde\bx_2
       \end{pmatrix}\right)^{\top}\bA(u)(\bR_1(u),\bR_2(u))
       \begin{pmatrix}
         \tilde\bx_1 \\
         \tilde\bx_2
       \end{pmatrix}\\
    =&\left(\bR_1(u)\bx_1+\bR_2(u)\bx_2\right)^{\top}\bA(u)(\bR_1(u)\bx_1+\bR_2(u)\bx_2) \\
    =& (\bR_1(u)\tilde\bx_1)^{\top}\bA(u)\bR_1(u)\tilde\bx_1 +(\bR_1(u)\tilde\bx_1)^{\top}\bA(u)\bR_2(u)\tilde\bx_2\\
    &+(\bR_2(u)\tilde\bx_2)^{\top}\bA(u)\bR_1(u)\tilde\bx_1+(\bR_2(u)\tilde\bx_2)^{\top}\bA(u)\bR_2(u)\tilde\bx_2\\
    =& \tilde\bx_1^{\top}(\bR_1(u))^{\top}\bA(u)\bR_1(u)\tilde\bx_1.
  \end{align*}
  The last equality holds because $\bA(u)\bR_2(u)=0$. Similarly, we
  can verify $2\bx^{\top}\bB(u)=2\tilde\bx_1\bR^{\top}_1\bB(u)$.
  Consequently, the optimal discriminant quadratic function is
  \[\tilde\bx_1^{\top}(\bR_1(u))^{\top}\bA(u)\bR_1(u)\tilde\bx_1+2\tilde\bx_1\bR^{\top}_1\bB(u)+C(u),\]
  which is independent of \(\tilde\bx_2\).
\end{myproof}

\putbib
\end{bibunit}

\end{document}